\documentclass[aps,pre,twocolumn,showpacs,amsmath,amssymb,superscriptaddress]{revtex4-1}
\usepackage{graphicx}
\usepackage{dcolumn}
\usepackage{bm}
\usepackage{natbib}
\usepackage{amsmath}
\usepackage{array}
\usepackage{hyperref}
\usepackage{cleveref}[2012/02/15]% v0.18.4; 
\crefformat{footnote}{#2\footnotemark[#1]#3}

\begin{document}
\title{Temperature-Dependent Defect Dynamics in the Network Glass SiO$_2$}

\author{Katharina Vollmayr-Lee}
 \email{kvollmay@bucknell.edu}
\affiliation{Department of Physics and Astronomy, Bucknell University,
      Lewisburg, Pennsylvania 17837, USA}
\affiliation{Georg-August-Universit\"at G\"ottingen,
Institut f\"ur Theoretische Physik,
Friedrich-Hund-Platz 1, 37077 G\"ottingen, Germany}
\author{Annette Zippelius}
\affiliation{Georg-August-Universit\"at G\"ottingen,
Institut f\"ur Theoretische Physik,
Friedrich-Hund-Platz 1, 37077 G\"ottingen, Germany}
\affiliation{Max-Planck-Institut f\"ur Dynamik und Selbstorganisation, Bunsenstr. 10,
37073 G\"ottingen, Germany}

\date{July 8, 2013}
%\date{\today}
\begin{abstract}

  We investigate the long time dynamics of a strong glass former,
  SiO$_2$, below the glass transition temperature % ($T_{\rm
   %c}=3330$~K)
  by averaging single particle trajectories over time windows which
  comprise roughly 100 particle oscillations. The structure on this
  coarse-grained time scale is very well defined in terms of
  coordination numbers, allowing us to identify ill-coordinated atoms,
  called defects in the following. The most numerous defects are 
  OO neighbors, whose lifetimes are comparable to the equilibration time
  at low temperature. On the other hand SiO and OSi defects are very rare 
  and short lived. The lifetime of defects is found to be strongly
  temperature dependent, consistent with activated processes. 
  Single-particle jumps give rise to local structural rearrangements. We
  show that in SiO$_2$ these structural rearrangements are coupled to
  the creation or annihilation of defects, giving rise to very strong
  correlations of jumping atoms and defects.
\end{abstract}

%http://www.aip.org/pacs/pacs2010/individuals/pacs2010_regular_edition/index.html
%%61.43.-j       Disordered solids
%%64.60.Bd       General theory of phase transitions
%%64.70.kj       Glasses
%%64.70.P-       Glass transitions of specific systems     (Heuer,Schroeder)
%%64.70.Q-       Theory and modeling of the glass transition
%45.70.-n 	Granular systems (see also 05.65.+b Self-organized systems)
%64.60.-i 	General studies of phase transition
%  64.70.kj 	Glasses 
%02.70.Ns, %}{Molecular dynamics and particle methods}
%{Glasses}
\maketitle
%\cite{colin2011,*HeuerJPCM2008,*MasriPRE2010,*parsaeian09,*Rehwald2010,*WarrenPRE2008,kvl2010}. 

\section{Introduction}

Amorphous SiO$_2$, or silica, has many fascinating features. Silica is of
importance in geology, chemistry, physics and industrial 
applications.
To classify the huge variety of glass formers in 
general \cite{glassbook,Ediger2000,AngellScience1995},
we distinguish fragile and strong glass 
formers \cite{Angellplot1976,Angell1991,AngellScience1995}.
Silica is a typical strong glass former, i.e. the shear viscosity 
exhibits Arrhenius behavior at low temperature and pressure.
With increasing temperature SiO$_2$ undergoes at $T_{\rm c}$ 
a strong to 
fragile transition \cite{BKSTc}
and for large pressure 
critical behavior of a liquid-liquid transition has been 
observed \cite{AngellHemmati2013}.

We investigate here SiO$_2$ via molecular dynamics 
simulations  using the van
Beest-Kramer-van Santen (BKS) potential 
\cite{beest_90} for the particle interactions.
Since previous simulations had shown that the BKS-potential is 
a very good model for real silica 
(\cite{vollmayr96_2,BKSTc,Badro1997,Taraskin1999}
and references therein) many simulations with the BKS-potential
followed, giving us insight into 
the phase diagram \cite{poole04,poole01,Badro1998,BarratBadro1997}, 
energy landscape \cite{saksaengwijit2004,reinisch2005,saksaengwijit2006,saksaengwijit2007,reinisch2006}, 
specific heat \cite{scheidler2001}, vibrational spectrum 
\cite{Taraskin1997,Taraskin1999,Taraskin2002,UchinoTaraskin2005,Leonforte2011},
dynamic heterogeneities \cite{Vogel2004,Bergroth2005,Teboul2006,Hung2013}, 
and aging  \cite{berthier07,kvl2010,kvl2013} 
\footnote{This is not  a complete list of BKS-simulations. 
For further work please see references therein.}.
%missing vink03

For temperatures below $T_{\rm c}=3330$~K \cite{BKSTc}, BKS-SiO$_2$ is
a strong glass former. A striking similarity with fragile glass
formers has been found for single particle jump dynamics
\cite{kvl2013}. This is surprising at first sight, because the local
structures in fragile and strong glasses differ considerably. In
fragile glasses the concept of a cage is well established and jumps
are interpreted as particles escaping from their cage. The underlying
structure in SiO$_2$, on the other hand, is based on randomly connected
tetrahedra, forming a macroscopic random network. Even though the
macroscopic network is random, coordination numbers are very well
defined, so that defects are easily identified. In this paper we
address the question to what extent jump events are correlated with
the creation of defects.

\begin{figure}[h]
\includegraphics[width=3.1in]{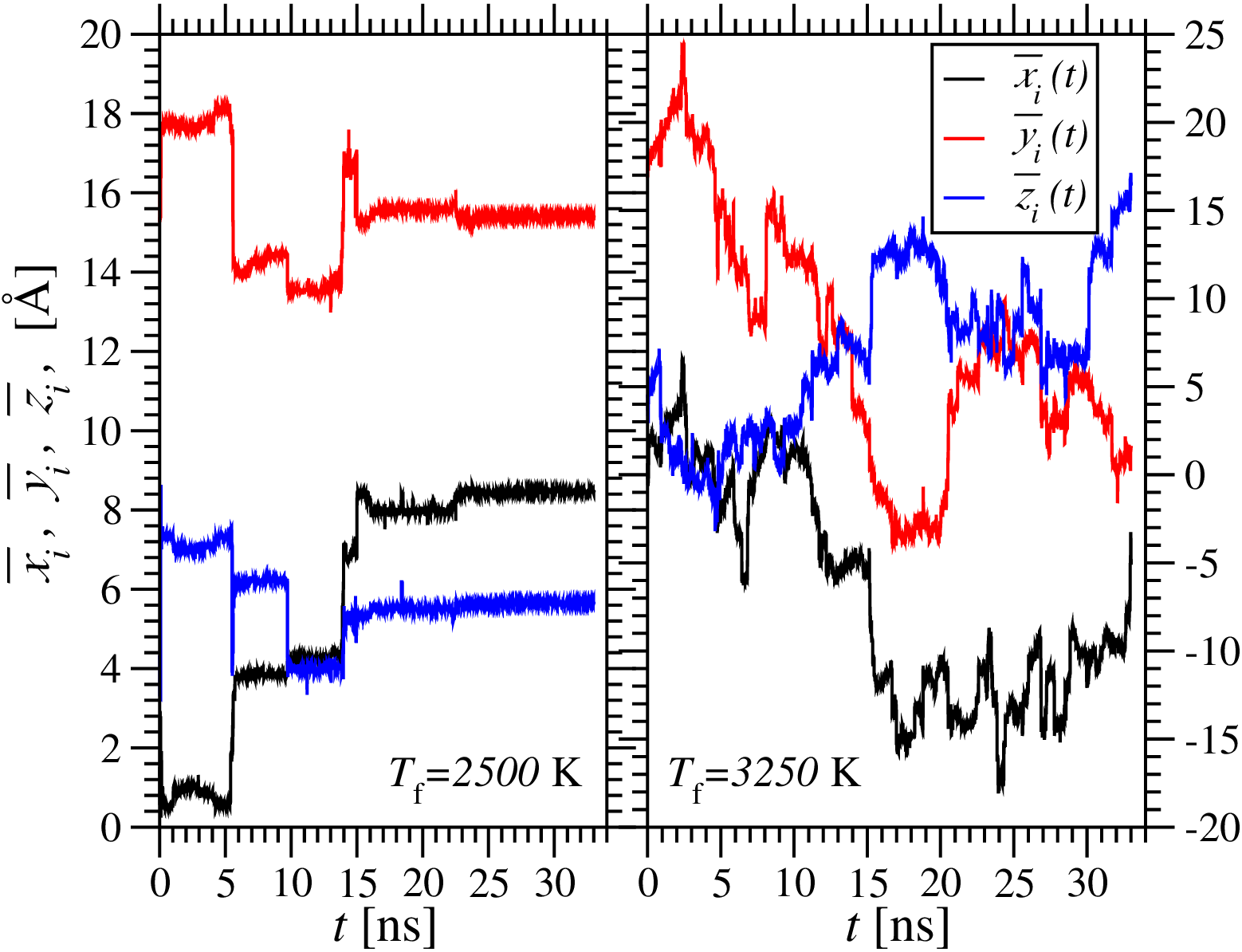}
\caption{\sf (color online) 
  Examples for time-averaged single particle (O-atom) trajectories 
  for $T_{\rm i}=5000$~K at $T_{\rm f}=2500$~K (left figure)
  and at $T_{\rm f}=3250$~K (right figure).
    }
\label{fig:traj_from5000c08p134}
\end{figure}

Our focus is on structural rearrangements well inside
the glassy phase. At low temperature, we expect a clear separation of
time scales, such that oscillations around preferred positions are
characterized by short time scales, whereas (rare) structural changes
occur on much longer time scales. To illuminate the latter, we filter
out short time oscillations by averaging particle trajectories and
analyze the {\em time-averaged single particle
  $i$ trajectories} $\overline{\mathbf r_i}(t)$ in terms of jumps of particles
and creation and annihilation of defects.  We find a clear temperature
dependence of the time-averaged dynamics which is already apparent in
single particle trajectories; an example is shown in
Fig.~\ref{fig:traj_from5000c08p134}. Whereas for the lowest
temperatures under consideration jump events are well separated in
time by long quiescent periods, this separation of time scales is
gradually lost, when the glass transition is approached from below.

Our approach is similar in spirit to the analysis of inherent
structures
\cite{Stillinger1984,Sastry1999,saksaengwijit2004,reinisch2005,saksaengwijit2006,saksaengwijit2007},
where instantaneous configurations are quenched to their local
potential energy minimum. In \cite{saksaengwijit2004,saksaengwijit2006}  
energy minimized configurations have been analyzed for BKS-SiO$_2$  
in order to explain the observed crossover from strong to fragile behavior. 
 We expect the time-averaged trajectories to
be strongly correlated with the corresponding inherent structures. 
However, in contrast to inherent structures, our approach allows us to
study the dynamics of local structural rearrangements.
Time averaged trajectories have been studied previously in refs.
\cite{OligschlegerPRB1999,kluge2006} for soft sphere
glasses and in \cite{Keys2011} for Lennard-Jones glasses. Keys et al.
\cite{Keys2011} use time persistent particle displacements to identify those 
excitations which are responsible for structural relaxation.
%earlierversion:
%Whereas time averaged trajectories have been studied previously 
%to detect jump 
%events \cite{kvl2013,Keys2011,Candelier2009,OligschlegerPRB1999,kluge2006}, 
%we use here time averaged trajectories also to enhance the 
%underlying local structure and therewith to detect defects.
%This allows us to directly measure correlations of jumps and defects.

After introducing the model in section \ref{sec:simulation}, we show
in section \ref{sec:gofr} that the radial distribution functions of
the time-averaged trajectories are considerably sharpened as compared
to the corresponding distributions for the unaveraged
trajectories. This implies a stable, well-defined structure on
time scales large compared to a typical oscillation period. 
This time-averaged structure is only weakly temperature dependent. 
The dynamics on intermediate and long time scales is dominated
by defects in the random network structure which are well 
defined in terms of the coordination number due to the sharp 
peak structure of the radial distribution function.
Number and lifetime of the defects are strongly temperature dependent
as discussed in section \ref{sec:defects}.  In section
\ref{sec:jumpdefect} we show that the jump events as defined in
\cite{kvl2013} are strongly correlated with the defects in the random 
network. We summarize our
results and draw conclusions in \ref{sec:summary}.

\section{Model and Simulation Details}
\label{sec:simulation}

To model amorphous SiO$_2$ we used the BKS potential 
\cite{beest_90}. We carried out molecular dynamics (MD)
simulations with $N_{\rm Si}=112$ silica atoms and $N_{\rm O}=224$
oxygen atoms and at constant volume 
$V=\left ( 16.920468 \mbox{\AA} \right )^3$, 
which corresponds to a density of $\rho=2.323$ g/cm$^3$.

At $6000$~K we generated 20 independent configurations, which 
then were fully equilibrated at initial temperature 
$T_{\rm i} \in \{5000$~K, $3760$~K$\}$  followed by 
an instantaneous quench to lower temperatures 
$T_{\rm f} \in \{ 2500$\,K, $2750$\,K, $3000$\,K, $3250$\,K$\}$, 
i.e. to temperatures below $T_c=3330$~K.
Unique to our simulations is that we applied the
Nos\'e-Hoover temperature bath at $T_{\rm f}$ only 
for the first $0.327$~ns (NVT)
and then continued with constant energy (NVE) for $32.7$~ns to disturb 
the dynamics minimally.  As shown in \cite{kvl2010} we confirmed 
that $T_{\rm f}$ stayed constant. 
The MD time step was $1.02$~fs and $1.6$~fs during the (NVT)
and (NVE) run respectively.
For further details of the simulations 
we refer the reader to \cite{kvl2010}.

We analyzed the combined (NVT) and (NVE) simulation runs at $T_{\rm
  f}$.  Specifically, for this paper we focus on major structural
events by analyzing time-averaged single particle trajectories
$\overline{{\mathbf r}_i}(t_l)$ at times $t_l=l\Delta t_{\rm av}$.
The typical time scale of an oscillation is around $3 \times 10^{-14}$ s,
roughly twenty times the MD step. The time average is taken over $\Delta
t_{\rm av}$ which has to be chosen large as compared to 
the oscillation time and
sufficiently small to resolve structural rearrangements such as 
single particle jumps and the creation and annihilation of defects.
For most of the data presented below, we have used
$\Delta t_{\rm av}=3.27 10^{-12}$ s, allowing for $l=1,\ldots,10100$
points of the trajectory, but we have checked other values of $\Delta
t_{\rm av}$ as well (see below).

\section{Radial Distribution Function And Coordination Number}
\label{sec:gofr}

We first discuss the structural properties of our system on time scales
long compared to a typical oscillation period. To that end we first
compute the radial distribution functions for the time-averaged
trajectories and compare them to the corresponding quantities for the
unaveraged trajectories, representing the structure on microscopic
time scales. We then go on to discuss the temperature dependence of the
time-averaged structure and the distribution of coordination numbers.

\subsection{Radial distribution function of time-averaged trajectories}
To analyze the local structure implied by the time-averaged
trajectories we compute

\begin{equation}
g_{\alpha \beta}(r)= 
     \left \langle 
       \frac{V}{N_{\alpha} N_{\beta}}
            \sum \limits_{i=1}^{N_{\alpha}}
            \sum \limits_{j=1 \atop j\ne i}^{N_{\beta}}
           \delta(\left |{\mathbf r} \right |
                  - \left | \overline{\mathbf r}_{ij}(t_l) \right |
      \right \rangle
       \mbox{,}
\label{eq:gofr}
\end{equation}
where $\alpha,\beta \in \{$Si,O$\}$ 
(for the case of $\alpha=\beta$ the denominator is
$N_{\alpha}\left (N_{\alpha}-1 \right )$), $\overline{\mathbf
  r}_{ij}(t_l)$ is defined via the time-averaged trajectories
$\overline{\mathbf r}_{ij}(t_l)=\overline{\mathbf r}_i(t_l)
-\overline{\mathbf r}_j(t_l)$.  To increase statistics in all following
(unless otherwise specified) the ensemble average $\left \langle
  \ldots \right \rangle$ is obtained via an average over 20
independent simulation runs and an average over $1000$ consecutive 
times, $t_l$,
starting at a waiting time $t_{\rm
  w}=16.35$~ns. %The only exception is Fig.\ref{fig:tauDDtw}, where the dependence on waiting time is tested.
For all following figures we used $T_{\rm i}=5000$~K. 
(This choice of $t_{\rm w}$ results from previous work \cite{kvl2013}; 
we have checked other waiting times as well as different $T_i$, see
\ref{sec:lifetime}). As will be shown below, the typical relaxation
times in our system are larger than $\Delta t_{\rm av}$, so that
configurations at different $t_l$ are not completely uncorrelated.
Therefore we determine error bars via the 20 independent simulation runs.

%-------------
\begin{figure}[h]
\includegraphics[width=3.1in]{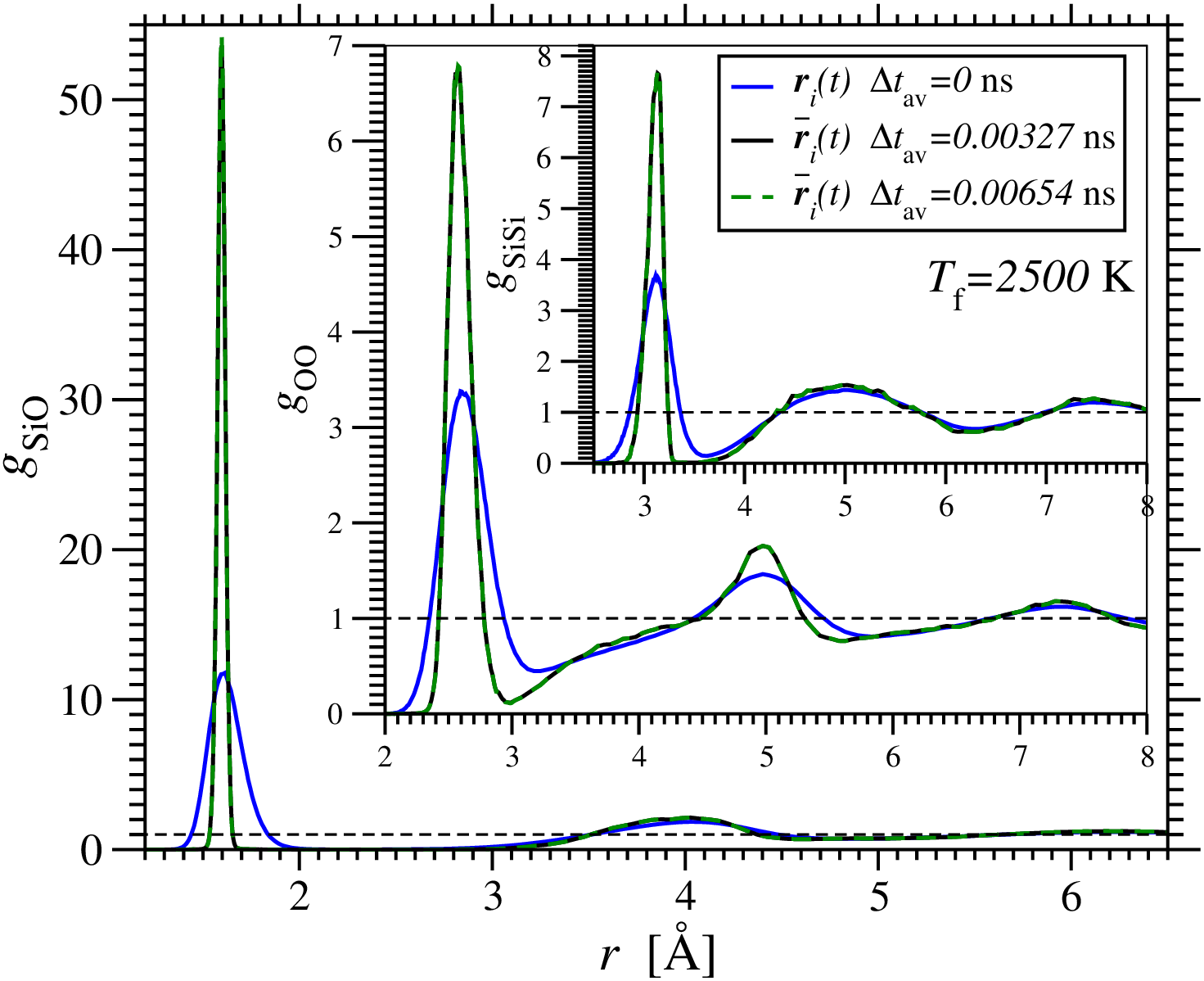}
\caption{\sf Radial distribution function $g_{\alpha \beta}(r)$ 
as defined in Eq.~(\ref{eq:gofr}) using different
time averages $\Delta t_{\rm av}$ for the time average of
${\overline {\mathbf r}_i}(t)$.
Here for final temperatures $T_{\rm f}=2500$~K quenched 
from $T_{\rm i}=5000$~K.}
%(additional info: $t_{\rm w}=16.35$~ns which is tstart5000, $N_{t_0}=1000$)
\label{fig:gofravcomp_T2500}
\end{figure}
%-------------
\begin{figure}[h]
\includegraphics[width=3.1in]{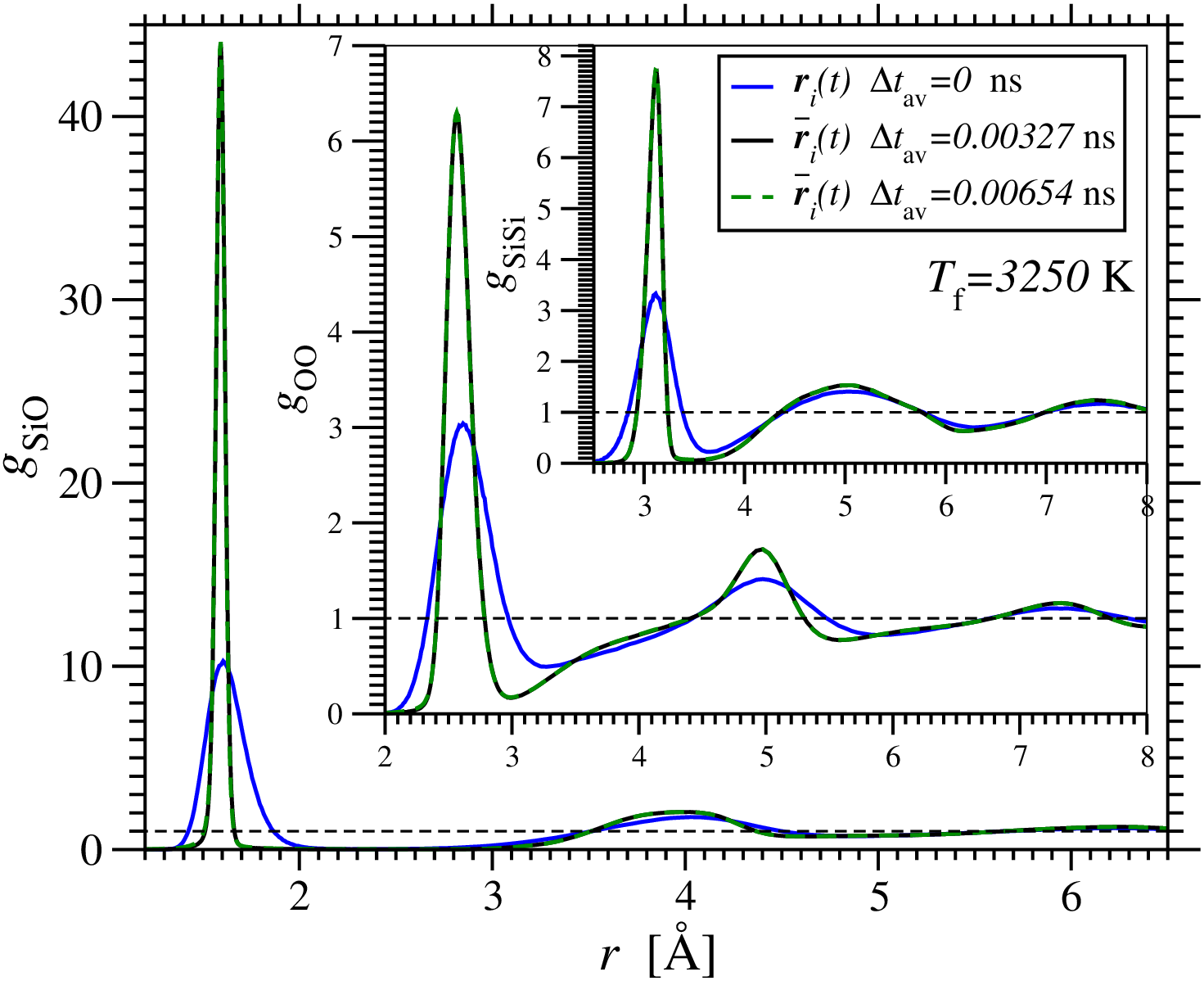}
\caption{\sf Radial distribution function as in 
     Fig.~\ref{fig:gofravcomp_T2500} but here for final temperature 
     $T_{\rm f}=3250$~K.
        }
\label{fig:gofravcomp_T3250}
\end{figure}

In Figs.~\ref{fig:gofravcomp_T2500} \& \ref{fig:gofravcomp_T3250} we
compare the pair correlation for trajectories with and without time
averaging.  We conclude that time averaging sharpens the pair
correlation drastically, both for low ($T_{\rm f}=2500$~K) and high
temperatures ($T_{\rm f}=3250$~K). The enhancement is particularly
strong for the nearest-neighbor peak of $g_{\rm SiO}$, implying that the
structural unit of one tetrahedron with an Si-atom at the center and
four O-atoms at the corners is well defined. We obtain the same 
$g_{\alpha \beta}(r)$ 
for $\Delta t_{\rm av}=0.00327$~ns and $\Delta t_{\rm av}=0.00654$~ns,
supporting the separation of time scales, so that time averaging
over $\Delta t_{\rm av}$ allows us to filter out the main structural
features of an underlying network that is highly ordered in the sense
that nearest neighbor distances are well-defined on time scales of the
order of $0.005$~ns.

\subsection{Temperature dependence of the time-averaged structure}

\begin{figure}[h]
\includegraphics[width=3.1in]{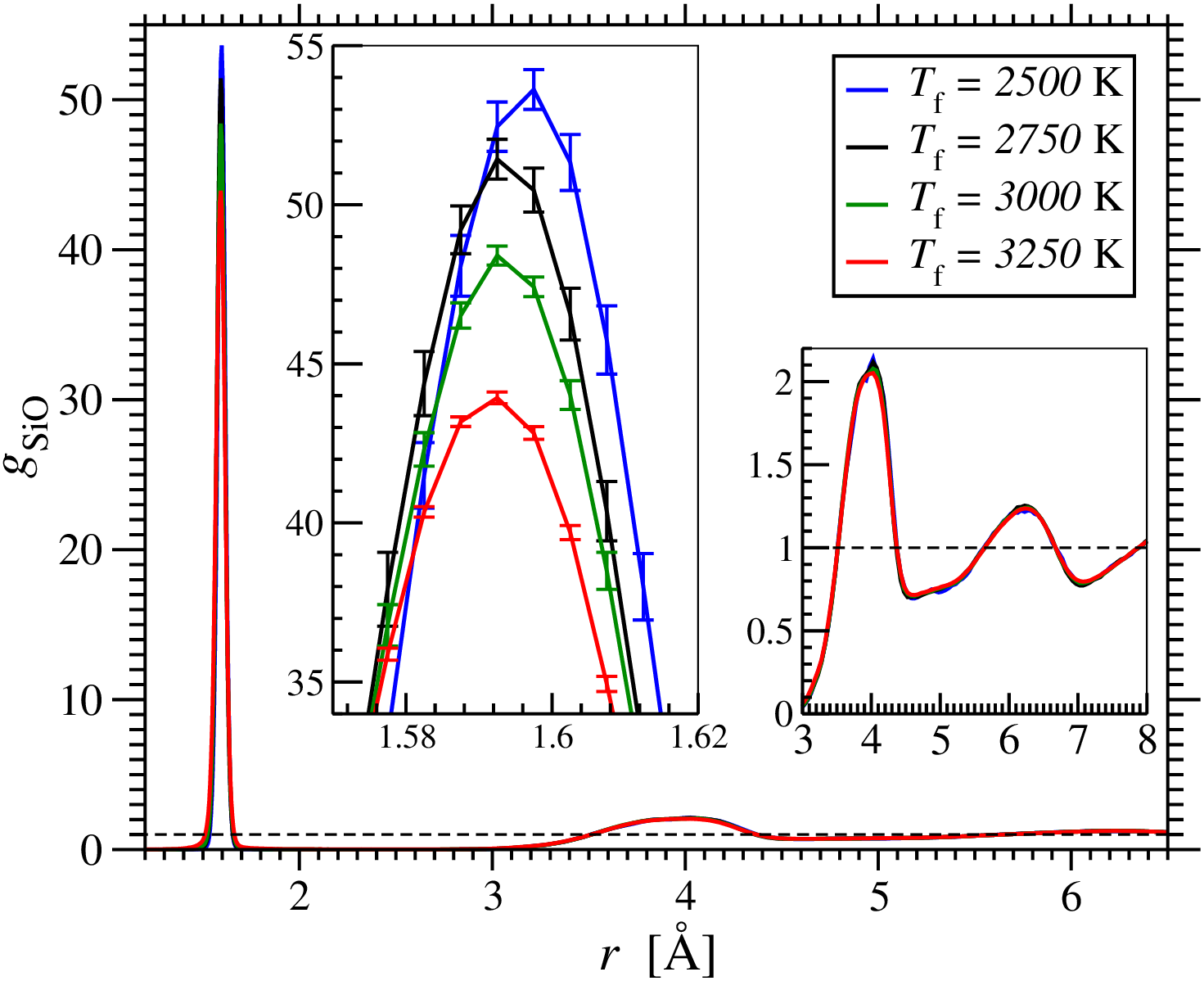}
\caption{\sf (color online) Radial distribution function 
$g_{\rm SiO}(r)$ for 
 final temperatures $T_{\rm f}$. % with $T_{\rm i}=5000$~K.
 The left inset is an enlargement of the first peak 
 and the right inset is an enlargement of farther peaks.
 %Error bars were determined via the 20 independent simulation runs. 
 %%are of the order of 0.6 for the first peak and 0.02 for 
 %%larger distances.
}
%(additional info: $t_{\rm w}=16.35$~ns which is tstart5000, $N_{t_0}=1000$)
\label{fig:gofr_SiO}
\end{figure}

\begin{figure}
\includegraphics[width=3.1in]{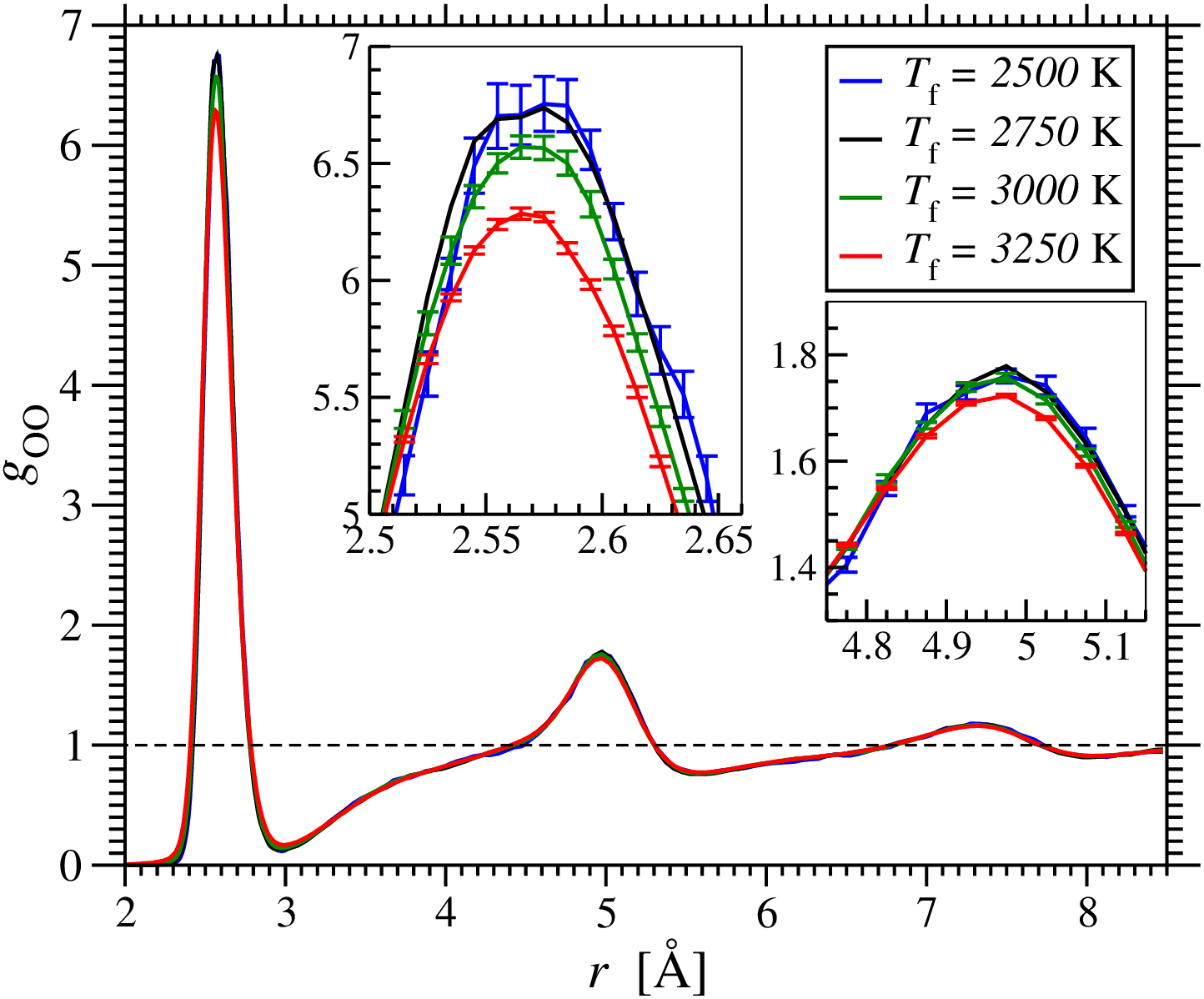}
\caption{\sf (color online) Similar to Fig.~\ref{fig:gofr_SiO} the
radial distribution function, but here for $g_{\rm OO}(r)$.
The insets are enlargements of the first and second peak.}
%Error bars about 0.1 for the first peak and 0.01 larger distances $r$.
%(additional info: $t_{\rm w}=16.35$~ns which is tstart5000, $N_{t_0}=1000$)
\label{fig:gofr_OO}
\end{figure}

Next we investigate, how the time-averaged structure depends on
temperature.  In Figs.~\ref{fig:gofr_SiO} and \ref{fig:gofr_OO} we
compare the radial distribution function, as defined in
Eq.~(\ref{eq:gofr}), for four different temperatures. For $g_{\rm
  SiO}(r)$ (see Fig.~\ref{fig:gofr_SiO}) we find an increase of roughly
$20\%$ for the temperature range investigated. Compared to the
increase by a factor of $\sim 5$ due to time averaging this is a
rather mild effect. Similarly, for $g_{\rm OO}(r)$ (see
Fig.~\ref{fig:gofr_OO}) the first peak is enhanced by roughly $5\%$, again
small as compared to the increase by a factor of $\sim 2$ due to time
averaging. For distances beyond nearest neighbors, the radial
distribution function of the time-averaged configurations is basically
temperature independent. We conclude that the time-averaged structure
is only weakly temperature dependent.
%-------------
%-------------
%\begin{figure}[h]
%\includegraphics[width=3.1in]{./figsv6.dir/gofr_SiSi_v2.eps}
%\caption{\sf (color online) Similar to Fig.~\ref{fig:gofr_SiO} the
%radial distribution function, but here for $g_{\rm SiSi}(r)$.}
%%(additional info: $t_{\rm w}=16.35$~ns which is tstart5000, $N_{t_0}=1000$)
%\label{fig:gofr_SiSi}
%\end{figure}
%-------------

\subsection{Coordination numbers}

\begin{figure}[h]
\includegraphics[width=3.1in]{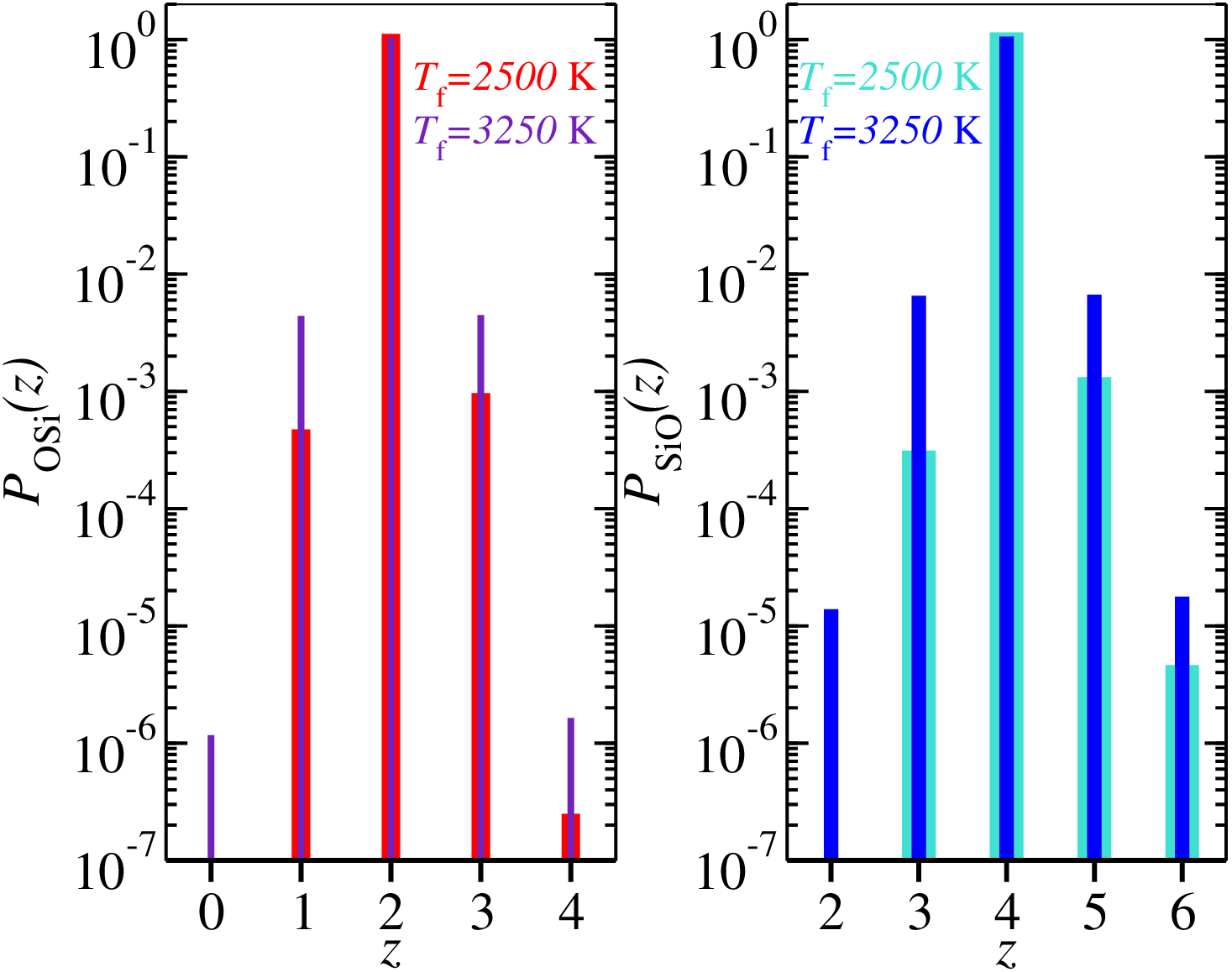}
\caption{\sf (color online) Distribution of coordination number $P(z)$ 
for the number of O-neighbors of an Si-atom ($P_{\rm SiO}$ in left panel) 
and for the number of Si-neighbors of an O-atom ($P_{\rm OSi}$ 
in right panel). Thick lines are for $T_{\rm f}=2500$~K and 
thin dark lines are for $T_{\rm f}=3250$~K.}
\label{fig:PofzSiOOSi}
\end{figure}

\begin{figure}[h]
\includegraphics[width=3.1in]{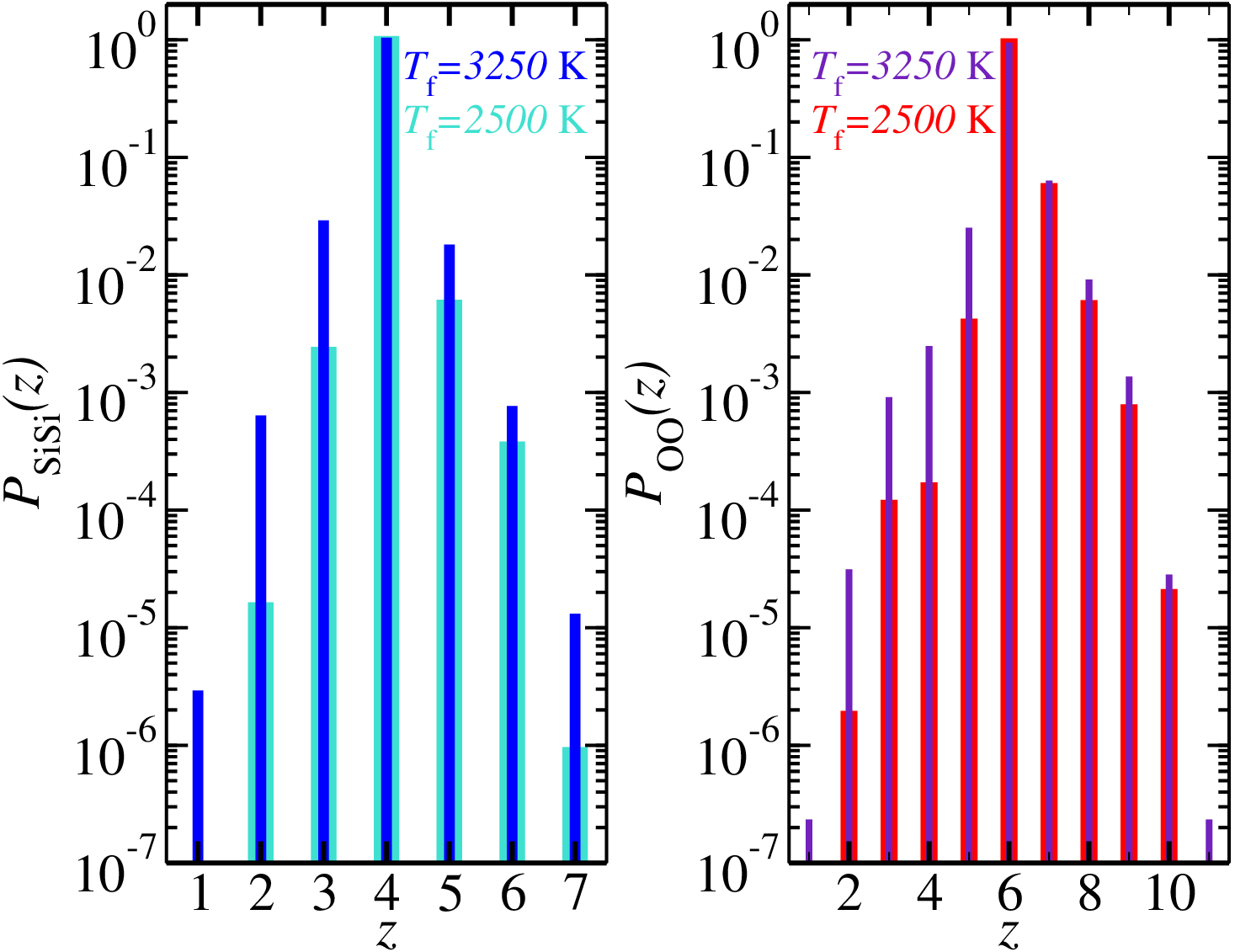}
\caption{\sf (color online) Distribution of coordination number $P(z)$ 
for the number of Si-neighbors of an Si-atom ($P_{\rm SiSi}$ in left
panel) and for the number of O-neighbors of an O-atom ($P_{\rm OO}$ 
in right panel). Thick lines are for $T_{\rm f}=2500$~K and 
thin dark lines are for $T_{\rm f}=3250$~K.}
\label{fig:PofzSiSiOO}
\end{figure}

For all investigated time-averaged $g_{\alpha \beta}(r)$
the first peak is very sharp and the minimum between 
the first and second peaks is very deep, indicating a 
well-defined first neighbor
shell. We therefore define for each particle $i$ of 
particle type $\alpha$ at time $t_l$ 
the coordination number $z_i^{\alpha \beta}(t_l)$ 
to be the number of other particles $j$ of type $\beta$ which
satisfy 

\begin{equation}
\left | \overline{\mathbf r}_i(t_l) 
  - \overline{\mathbf r}_j(t_l) \right | < r_{\rm min}^{\alpha \beta} 
\label{eq:zi}
\end{equation}
where $r_{\rm min}^{\rm SiSi}=3.42$~\AA, $r_{\rm min}^{\rm
  SiO}=2.40$~\AA $\,$ and $r_{\rm min}^{\rm OO}=3.00$~\AA.  The
resulting coordination number distributions $P_{\alpha \beta}$ are
plotted in Figs.~\ref{fig:PofzSiOOSi} \& \ref{fig:PofzSiSiOO},
ensemble averaged as in Eq.~(\ref{eq:gofr}).

%-------------
%-------------
%-------------
% and forms a network 
%of 0-corner sharing SiO$_4$-tetrahedra. 

Please note that the distributions $P(z)$ are so sharply peaked that
we chose a logarithmic scale. At $T_{\rm f}=2500$~K (thick lines)
$99.9$\% of Si-atoms are surrounded by four O-atoms and $99.9$\% of
O-atoms are surrounded by two Si-atoms.  Even at $T_{\rm f}=3250$~K
(thin dark lines) there are $98.8$\% Si-atoms with $z_i^{\rm SiO}=4$
and $99.1$\% O-atoms with $z_i^{\rm OSi}=2$.  The time-averaged
configurations form an almost perfect O-corner sharing network of
SiO$_4$-tetrahedra.  $P_{\rm SiSi}(z)$ probes this network on the
length scale of tetrahedra to tetrahedra connections.  Also on this
length scale we find that the coordination is almost perfect in the
time-averaged configurations: for $T_{\rm f}=3250$~K ($T_{\rm
  f}=2500$~K) $95.5$\% ($99.1$\%) of Si-atoms are surrounded by four
Si-atoms.  The broadest distribution is $P_{\rm OO}(z)$ for which
$90$\% O-atoms are surrounded by $z_i^{\rm OO}=6$ O-atoms.

\section{Defects}
\label{sec:defects}
In the previous section we showed that the 
time-averaged configurations form an almost perfect 
network with respect 
to the coordination number $z_i^{\alpha \beta}$
with $z^{\rm SiSi}_{\rm perfect}=4$, $z^{\rm Si0}_{\rm perfect}=4$,
 $z^{\rm OSi}_{\rm perfect}=2$ and $z^{\rm OO}_{\rm perfect}=6$.
For the rest of the paper we observe the dynamics of
the system by focusing not on this perfect structure, but 
rather on the deviations from it.

\subsection{Number of Defects}

\begin{figure}[h]
\includegraphics[width=3.1in]{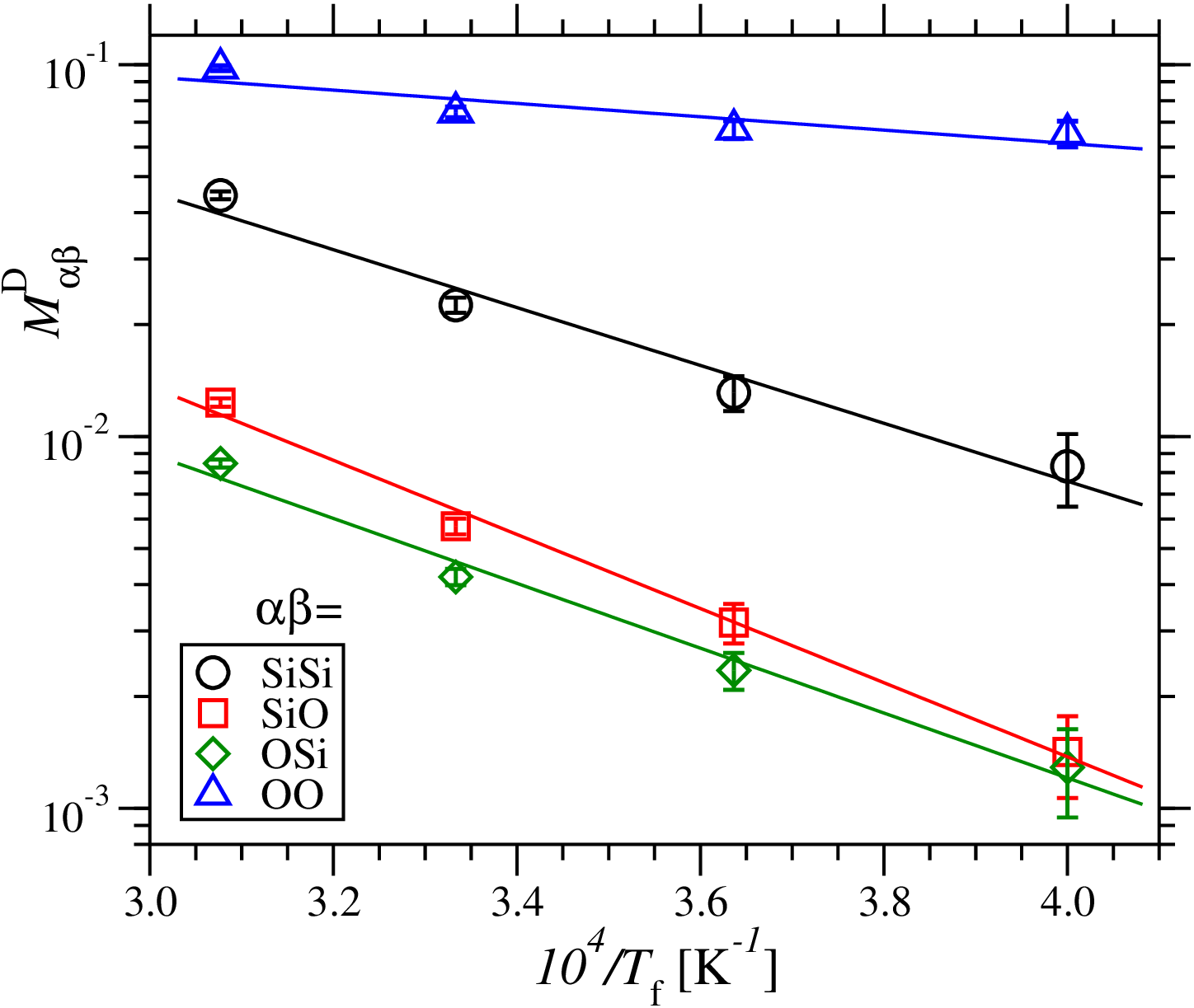}
\caption{\sf (color online) $M_{\alpha \beta}^{\rm D}$, 
 the fraction of particles $i$ of type $\alpha \in \{$Si,O$\}$
 which are defects, i.e. for which
  $z^{\alpha \beta} \ne z^{\alpha \beta}_{\rm perfect}$ (symbols) versus $1/ T_{\rm f}$;
also shown are Arrhenius fits 
  $M^{\rm D}_{\alpha \beta}(T_{\rm f})
      =C \exp \left (\frac{-E_{\rm A}}{k T_{\rm f}} \right )$ 
  (lines), with $C=9.8/13.4/3.7/0.3$ and 
  $E_{\rm A}=1.5$~eV$/2.0$~eV$/1.7$~eV$/0.36$~eV for 
  SiSi/SiO/OSi/OO-defects respectively.
}
%(additional info: $t_{\rm w}=16.35$~ns which is tstart5000, $N_{t_0}=1000$)
\label{fig:MDofT}
\end{figure}

We identify defects in the time-averaged structure with help of an
indicator function, defined for particle $i$ of type $\alpha$: 

\begin{equation}
      \chi^{\rm D}_i(t_l,\beta) = \left \{
            \begin{array}{ll}
              1 & \mbox{if at time $t_l$}\qquad z_i^{\alpha \beta} \ne z^{\alpha \beta}_{\rm perfect} \\
              0 & \mbox{if at time $t_l$}\qquad z_i^{\alpha \beta} = z^{\alpha \beta}_{\rm perfect} 
        \mbox{.}
            \end{array}
                                       \right . 
\label{eq:chiD}
\end{equation}
This means that an $\alpha \beta$-defect occurs if a particle of 
type $\alpha$ is surrounded by 
$z_i^{\alpha \beta} \ne z^{\alpha \beta}_{\rm perfect}$ particles of 
type $\beta$.
In Fig.~\ref{fig:MDofT} we show the fraction of particles which are defects, 

\begin{equation}
       M_{\alpha \beta}^{\rm D}=\left \langle 
         \frac{1}{N_{\alpha}} \sum \limits_{i=1}^{N_{\alpha}}
         \chi^{\rm D}_i(t_l,\beta) 
       \right \rangle \mbox{.}
\label{eq:MDofT}
\end{equation}

%-------------
\vspace*{5mm}

Consistent with the above coordination number distributions, $M_{\rm
  SiO}^{\rm D}$ and $M_{\rm OSi}^{\rm D}$ are very small. Most
defects are OO-defects. 
These findings give further support to the
picture of very stable tetrahedra
%being connected to each other in a more flexible way.
with relaxation processes mainly due
to rearrangements of the SiO$_4$-tetrahedra with respect to each other. 
%keeping mostly SiO-bonds in tact and intermittent states with a 
%broad SiOSi-bond angle distribution \cite{vollmayr96_2,BKSTc,Taraskin1997}.
With increasing temperature, the fraction of
defects increases approximately following Arrhenius behavior 
(with the exception of
$M_{\rm OO}^{\rm D}$, which is equally well fitted by a power law).
These results are in accordance with the work of Horbach and Kob 
\cite{BKSTc} who, however, use non-time-averaged configurations 
and therefore find more defects.

\subsection{Life Time of Defects via Time Correlation}
\label{sec:lifetime}

So far, we have simply counted the number of defects.
Next we look in more detail by observing the defects 
as they change with time. We ask the question whether 
the few defects are long lived defects of the same few particles
or if instead the defects are short lived, i.e. come and go 
over the simulation run at different locations.
To address this question we define a correlation 
function for defects of the same particle $i$ occurring 
at different times $t_l$ and $(t_l+t)$:

%\begin{widetext} %single column
\begin{align}
%C^{\rm DD}(t,\alpha,\beta,t_{\rm w}) = 
C^{\rm DD}(t,\alpha,\beta) = 
   \left \langle 
     \frac{1}{N_{\alpha}} \sum \limits_{i=1}^{N_{\alpha}}
   \chi^{\rm D}_i(t_l,\beta) \chi^{\rm D}_i(t_l+t,\beta) 
      \right \rangle\nonumber\\
    -
     \left \langle 
     \frac{1}{N_{\alpha}} \sum \limits_{i=1}^{N_{\alpha}}
     \chi^{\rm D}_i(t_l,\beta) 
      \right \rangle
     \left \langle 
     \frac{1}{N_{\alpha}} \sum \limits_{i=1}^{N_{\alpha}}
     \chi^{\rm D}_i(t_l+t,\beta) 
      \right \rangle
     \mbox{.}
\label{eq:CDDt0ptoft}
\end{align}
%\end{widetext} %single column
%The second term on the right side of Eq.~(\ref{eq:CDDt0ptoft}) ensures 
%that trivial correlations are excluded and therefore 
%$C^{\rm DD}(t \to \infty)$ goes to zero.
For the comparison of 
different $\alpha$, $\beta$ we normalize by the initial value, 

\begin{equation}
  \tilde{C}^{\rm DD}_{\alpha,\beta}(t)=
        \frac{C^{\rm DD}(t,\alpha,\beta)}
             {C^{\rm DD}(t=0,\alpha,\beta)}
     \mbox{.}
\label{eq:CDDtildeoft}
\end{equation}
%-------------

\begin{figure}[h]
\includegraphics[width=3.1in]{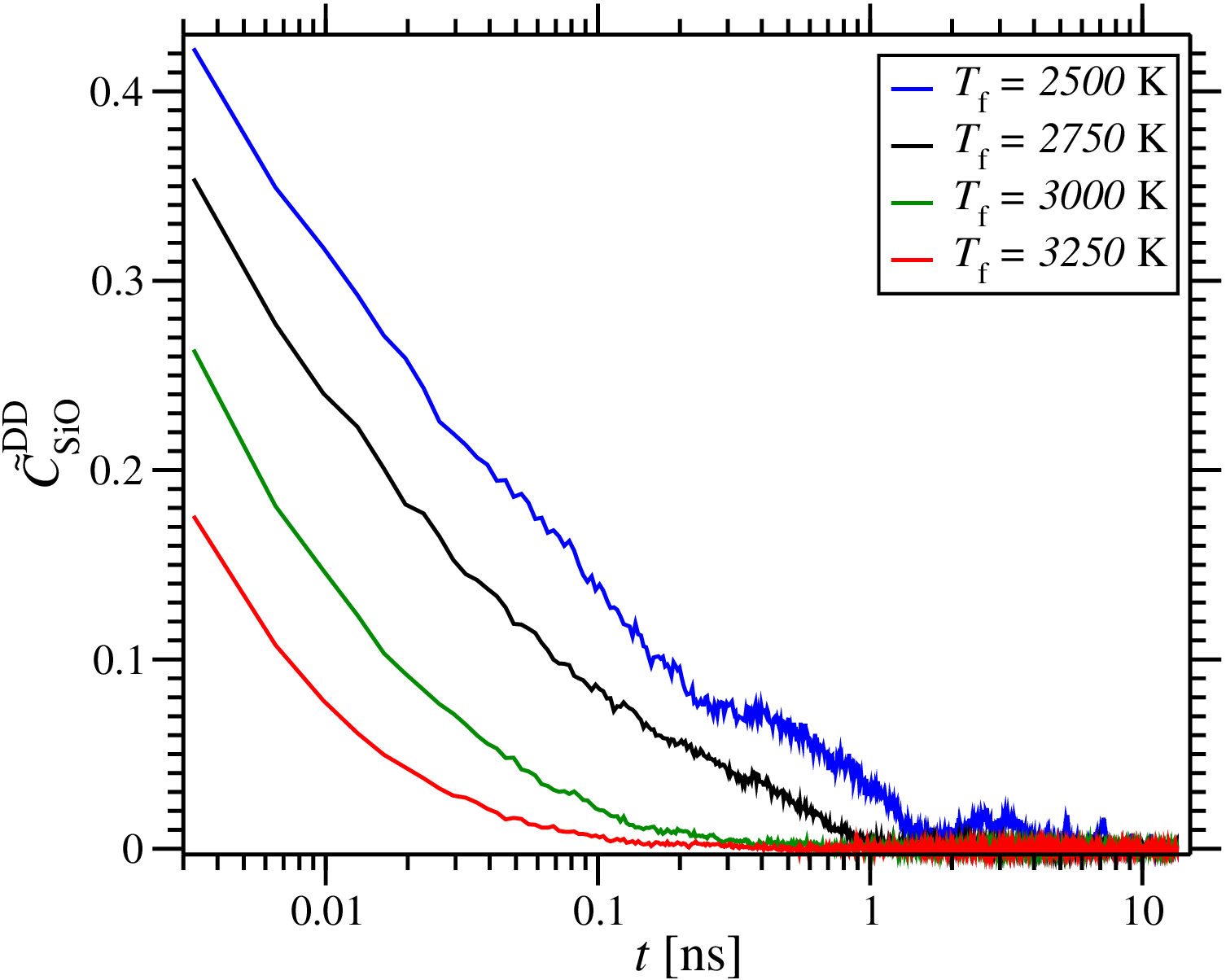}
\caption{\sf (color online) Normalized time correlation 
$\tilde{C}^{\rm DD}_{\rm SiO}(t)$ as defined in
Eqs.(\ref{eq:CDDt0ptoft}) \& (\ref{eq:CDDtildeoft}) for final 
temperatures $T_{\rm f}=2500$~K (top curve) to $T_{\rm f}=3250$~K bottom 
curve.}
%(additional info: $t_{\rm w}=16.35$~ns which is tstart5000, $N_{t_0}=1000$)
\label{fig:CDDoft_SiO}
\end{figure}

\begin{figure}[h]
\includegraphics[width=3.1in]{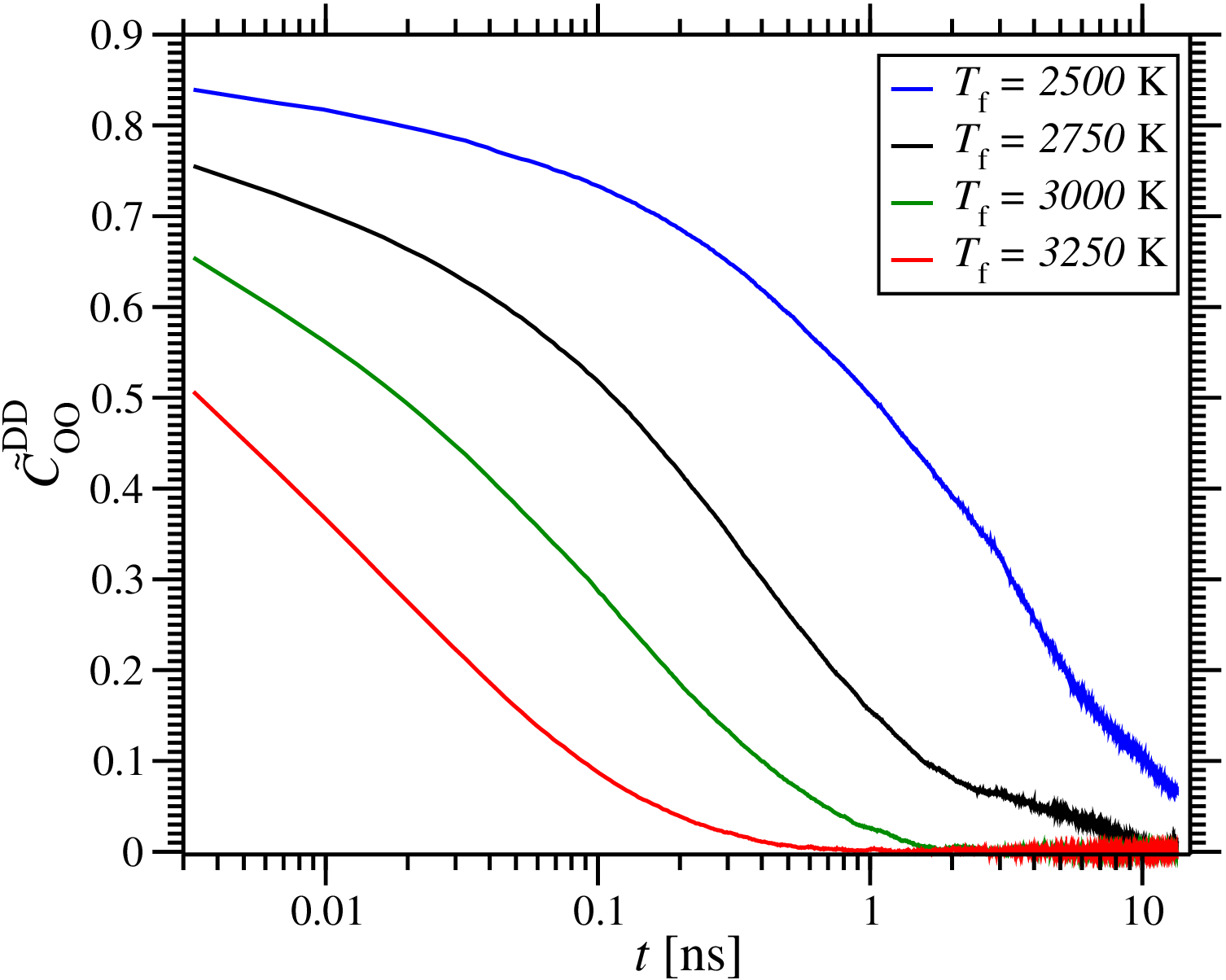}
\caption{\sf (color online) Similar to Fig.~\ref{fig:CDDoft_SiO}
but here for $\tilde{C}^{\rm DD}_{\rm OO}$.}
%(additional info: $t_{\rm w}=16.35$~ns which is tstart5000, $N_{t_0}=1000$)
\label{fig:CDDoft_OO}
\end{figure}
\begin{figure}[h]
\includegraphics[width=3.1in]{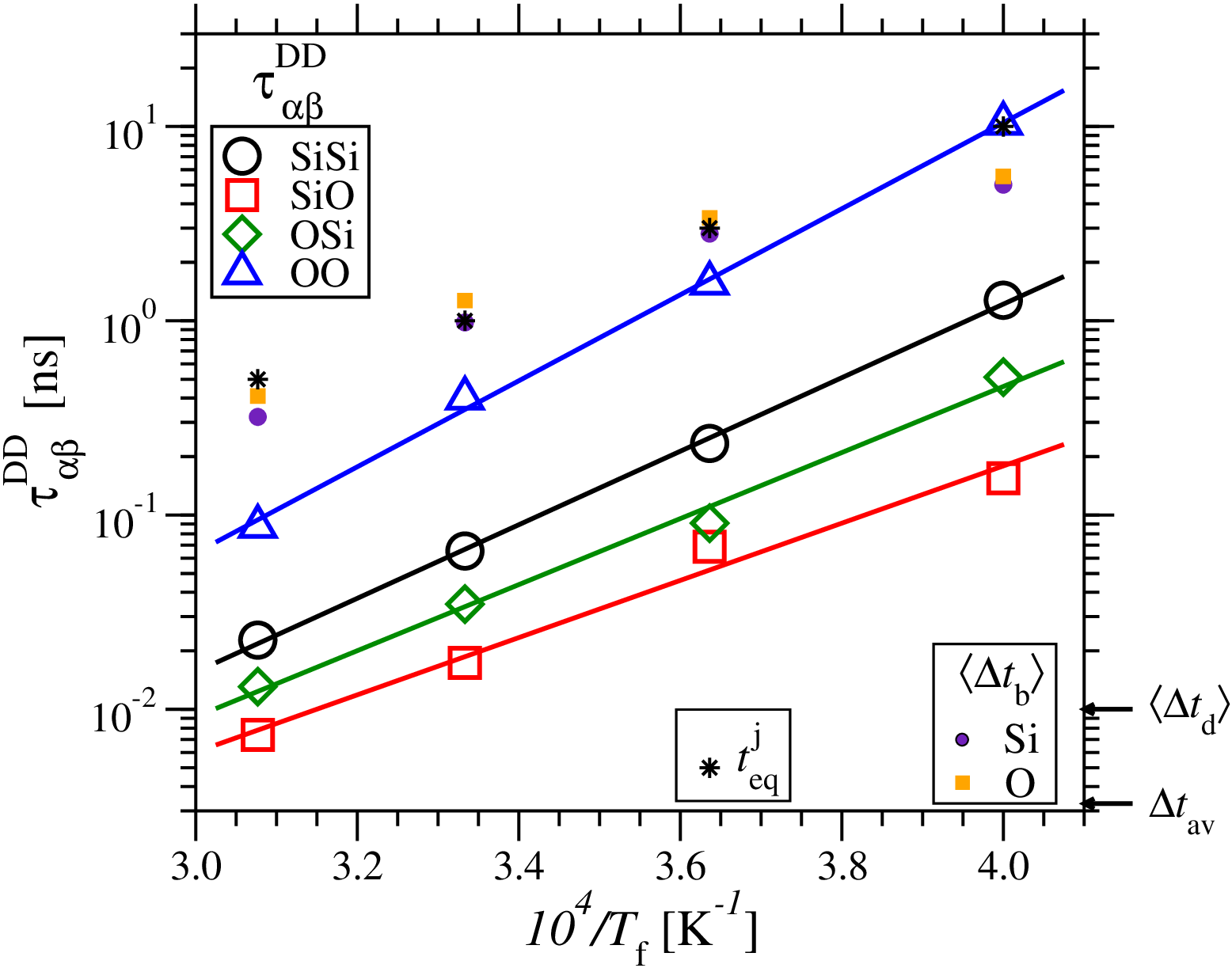}
\caption{\sf (color online) $\tau^{\rm DD}_{\alpha \beta}(10^4/T_{\rm f})$
 for various $\alpha,\beta \in \{$Si,O$\}$ and 
for fixed $T_{\rm i}=5000$~K and $t_{\rm w}=16.35$~ns (large open symbols).
Included are fits (lines) $\tau^{\rm DD}_{\alpha \beta}=C \exp \left( \frac{E_{\rm A}}{k T_{\rm f}} \right)$ with
$E_{\rm A}=3.76/2.92/3.37/4.39$~eV for SiSi/SiO/OSi/OO 
respectively. For comparison we show the time averaging interval 
$\Delta t_{\rm av}$ (lower arrow), the time duration of jumps 
$\langle \Delta t_{\rm d} \rangle$ (upper arrow), the time between
successive jumps $\langle \Delta t_{\rm b} \rangle$ (small filled symbols), 
and the equilibrium time $t_{\rm eq}^j$ (stars).
}
%(additional info: tstart5000, $N_{t_0}=1000$,t0pt)
\label{fig:tauDDofinvT}
\end{figure}

%%-------------
Figs.~\ref{fig:CDDoft_SiO} and \ref{fig:CDDoft_OO} reveal a strong
temperature dependence: with increasing temperature, all
$\tilde{C}^{\rm DD}_{\alpha \beta}$ decay faster. To quantify this
decay for various defect types $\alpha \beta$, % initial temperature
%$T_{\rm i}$, and waiting time $t_{\rm w}$,
we define the lifetime $\tau^{\rm DD}_{\alpha \beta}$ as the time
when $\tilde{C}^{\rm DD}_{\alpha,\beta}(\tau^{\rm DD}_{\alpha \beta})=0.1$  
(we find qualitatively the same results for
other values than $0.1$).  In Fig.~\ref{fig:tauDDofinvT} we show $\tau^{\rm
  DD}_{\alpha \beta}$ as a function of inverse temperature. For all
defect types $\alpha,\beta$ the defect lifetime $\tau^{\rm
  DD}_{\alpha \beta}(T_{\rm f})$ increases with decreasing $T_{\rm
  f}$. The increase is strongest for OO (a factor of 100), and weakest
for SiO (a factor of 20).

For $T_{\rm f} < T_{\rm c}$  Horbach and Kob \cite{BKSTc} and Saksaengwijit
and Heuer \cite{saksaengwijit2006} find  
Arrhenius behavior for the life time of SiO-bonds. We therefore
compare our data in Fig.~\ref{fig:tauDDofinvT} with Arrhenius fits
(lines). We find good agreement with excitation energies of the same
order as found in \cite{saksaengwijit2006}, although
refs.\cite{BKSTc,saksaengwijit2006} discuss the lifetime of bonds
instead of defects, as is done here.

One of the most startling observations in
 Figs.~\ref{fig:CDDoft_SiO},\ref{fig:CDDoft_OO} and \ref{fig:tauDDofinvT}
is the difference in lifetime for the various types of defects. 
To judge whether these life times are short or long, we
compare with the results of \cite{kvl2013} and \cite{kvl2010}.
Single particle (sudden) jump events are of the duration of 
$\langle \Delta t_{\rm d} \rangle = 0.01$~ns (see upper arrow 
in Fig.~\ref{fig:tauDDofinvT}) and the
average time spent between successive 
jumps $\langle \Delta t_{\rm b} \rangle$ is shown 
with filled small symbols.
% $= 6/4/1.2/0.4$~ns for $2500/2750/3000/3250$~K. 
This jump dynamics as well as the incoherent intermediate 
scattering function become waiting time
independent, i.e. the system reaches equilibrium, at 
$t_{\rm eq}^{\rm j} \approx t_{\rm eq}^C$ %$10/3/1/0.3$~ns.
(see stars in Fig.~\ref{fig:tauDDofinvT}).
%\\(additional info: $t_r^{C_q}$ via generalized incoherent intermediate
%        scattering function ($\alpha$-relaxation time) for
%        $q=1.7$~\AA is $3.35/1.15/0.17/0.10$~ns for
%        $2500/2750/3000/3250$~K \cite{kvl2010} \\
%         (for $T_{\rm f}=2500$~K indicate newest results with
%          Horacio C. that this time is even larger than $10$~ns).)

We conclude that SiO and OSi defects survive the duration of a jump
but are rather short lived when compared to the time span between
jumps; for all $T_{\rm f}$ the lifetime is less than $0.1 \times
\langle \Delta t_{\rm b} \rangle$ and for $T_{\rm f}=3250$~K the 
lifetime even becomes comparable to the duration of jump
events. OO-defects, on the other hand, are of the order of $t_{\rm
  eq}^{\rm j}$, implying that they are the excitations with the
longest lifetime in the system under study. Note that all defect
lifetimes are substantially longer than $\Delta t_{\rm av}$, which is
indicated by the lower arrow in Fig.\ref{fig:tauDDofinvT}.

We have checked that these results are robust with respect to choice of 
 $T_{\rm i}$ and waiting time $t_{\rm w}$. For $T_{\rm f}=2500$~K and $2750$~K
the results are qualitatively the same for 
$T_{\rm i}=3760$~K and $5000$~K and $0\leq t_{\rm w}\leq 26.2$ ns; 
for larger temperatures the results agree even quantitatively.

%-------------
\section{Correlation of Jumps and Defects}
\label{sec:jumpdefect}

Before discussing the correlations of jump events and defects, we
briefly recall the methods \cite{kvl2013} to identify and analyze jump
events.
\subsection{Jumps}
\begin{figure}[h]
\includegraphics[width=3.1in]{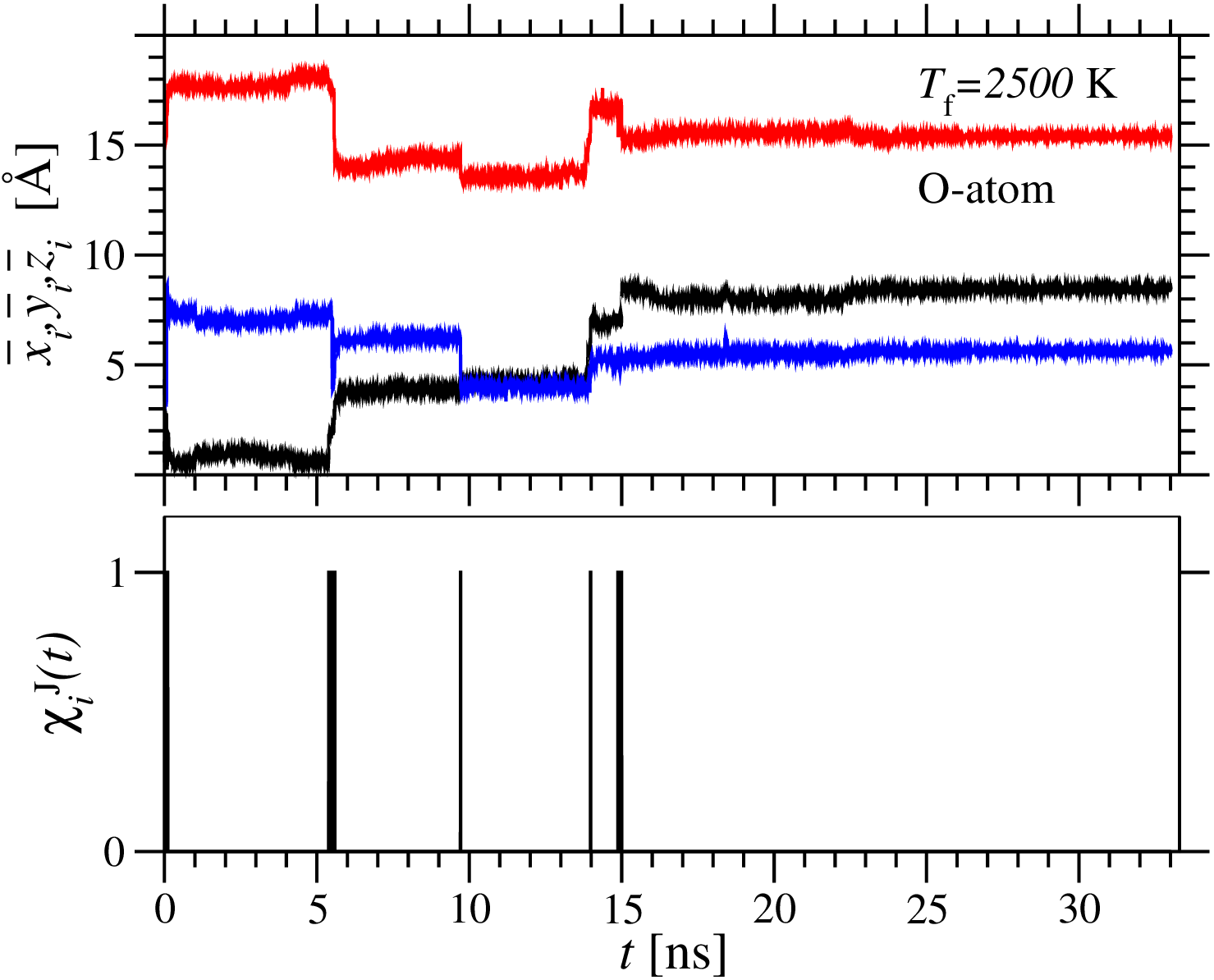}
\caption{\sf (color online) Time averaged trajectory of O-atom 
 at $T_{\rm i}=5000$~K (same as left panel of 
 Fig.~\ref{fig:traj_from5000c08p134}) 
 and $\chi_i^{\rm J}(t)$ which indicates jump events as horizontal lines.}
\label{fig:traj_chiJ_2500from5000c08p134}
\end{figure}

Whereas in the previous section %\ref{sec:defects} 
we characterized the relaxation 
dynamics with defects and their occurrence as function 
of time $\chi_i^{\rm D}(t_l,\beta)$, 
we now follow the approach of \cite{kvl2013}. For each 
time-averaged single-particle trajectory $\overline{\mathbf r}_i(t_l)$
we identify single-particle jump events using Eq.~(2) of \cite{kvl2013}. 
This means that a jump event of particle $i$ occurs  if 
\begin{equation}
 \left | \overline{\mathbf r}_i(t_l)-\overline{\mathbf r}_i(t_{l-4}) \right |
   > 3 \sigma_{\alpha}
\label{eq:jumpdef}
\end{equation}
holds, where $\sigma_{\alpha}$ is the average fluctuation size for
particle $i$ of type $\alpha$. Please note that all following results
are qualitatively the same if we use instead of the factor $3$ the
factor $\sqrt{2}$.  Numbering the jump events of particle $i$ by $k$, 
we determine for each jump event the time $t_k^{\rm init}$ when the
particle starts to jump and the time $t_k^{\rm f}$ when the particle
jump is finished.  We then define an indicator function for the
jumpers, in close analogy to the defects, by
\begin{equation}
      \chi^{\rm J}_i(t_l) = \left \{
            \begin{array}{ll}
              1 & \mbox{if}\quad t_k^{\rm init} \le t_l \le t_k^{\rm f}
                       \quad \mbox{for jumps $k$}\\
              0 & \mbox{otherwise} \mbox{.}
            \end{array}
                                       \right . 
\label{eq:chiJ}
\end{equation}
An example for the trajectory in the left panel of
Fig.~\ref{fig:traj_from5000c08p134} is shown in  Fig.~\ref{fig:traj_chiJ_2500from5000c08p134}, clearly revealing the jump events.

%-------------

%As illustrated in Fig.~\ref{fig:traj_chiJ_2500from5000c08p134}
%we now can characterize the relaxation dynamics by defining
%%Similar to $\chi_i^{\rm D}(t_l,\beta)$ 
%$\chi^{\rm J}_i(t_l)$ to be only nonzero during a jump
%\begin{equation}
%      \chi^{\rm J}_i(t_l) = \left \{
%            \begin{array}{ll}
%              1 & \mbox{if}\quad t_k^{\rm init} \le t_l \le t_k^{\rm f}
%                       \quad \mbox{for jumps $k_i$}\\
%              0 & \mbox{otherwise}
%            \end{array}
%                                       \right . 
%\label{eq:chiJ}
%\end{equation}
 
\begin{figure}[h]
\includegraphics[width=3.1in]{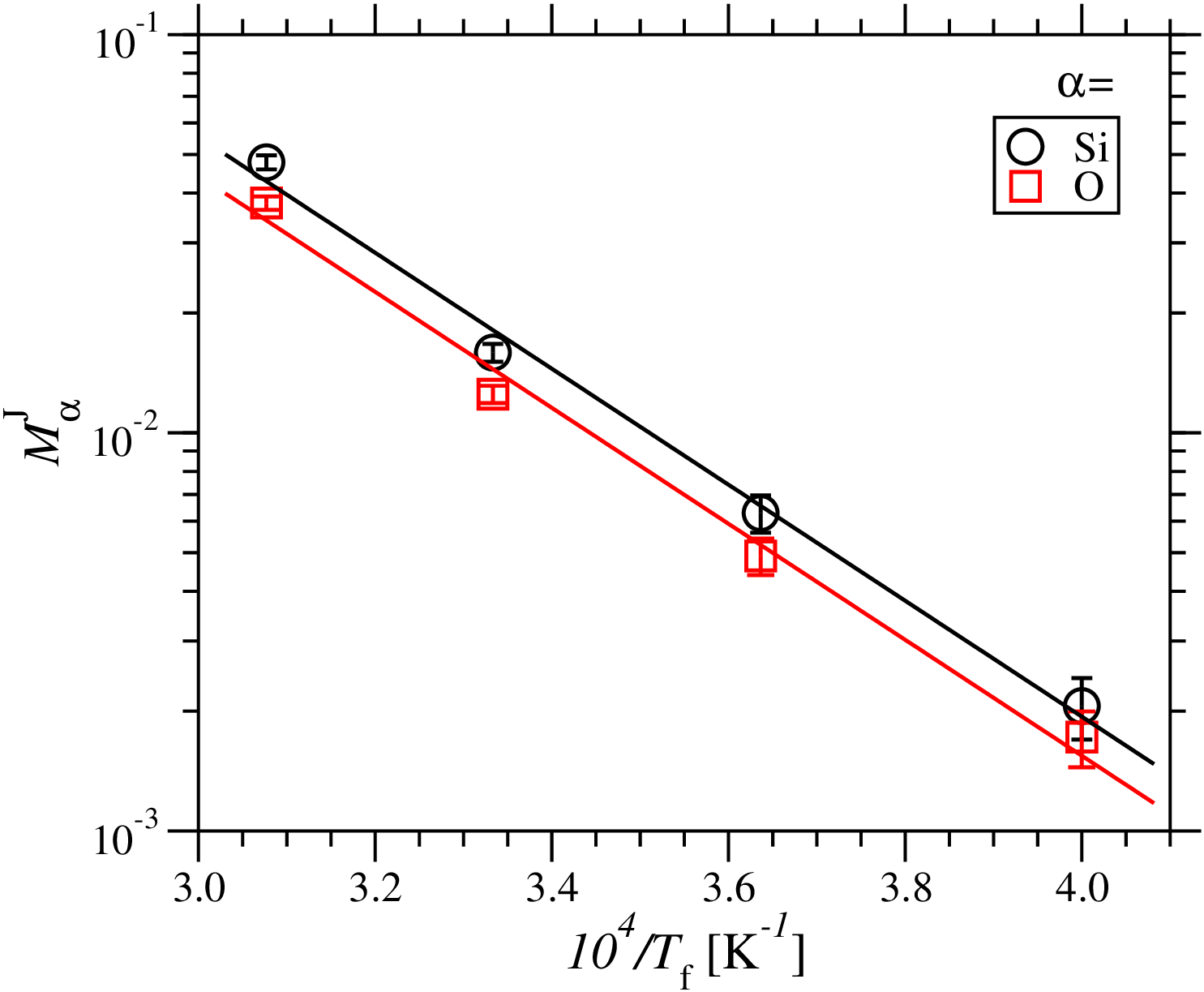}
\caption{\sf (color online) The fraction of jumping particles 
$M^{\rm J}_{\alpha}$ (symbols) versus $1/T_{\rm f}$ and fits
with 
%$M^{\rm J}_{\alpha}(T_{\rm f})=C \exp \left (\frac{k T_{\rm f}}{E_{\rm A}} \right )$ (solid lines) with $E_{\rm A}=0.0208$~eV \& $E_{\rm A}=0.0212$~eV 
%for $\alpha=$Si \& $\alpha=$O,
$M^{\rm J}_{\alpha}(T_{\rm f})=C \exp \left (\frac{-E_{\rm A}}{k T_{\rm f}} \right )$ (lines) with $E_{\rm A}=2.89$~eV.}
%and $M^{\rm J}_{\alpha}(T_{\rm f})=C \left ( T_{\rm f} \right )^{\nu}$ 
%(dashed lines) with $\nu=11.8$ \& $11.6$ for $\alpha=$Si \& $\alpha=$O},
\label{fig:MJofTfits}
\end{figure}

%and we obtain  instead of $M^{\rm D}_{\alpha \beta}$  now 
The average fraction of jumping particles, 

\begin{equation}
       M^{\rm J}_{\alpha}=\left \langle 
         \frac{1}{N_{\alpha}} \sum \limits_{i=1}^{N_{\alpha}}
         \chi^{\rm J}_i(t_l) 
       \right \rangle
       \mbox{,}
\label{eq:MJofT}
\end{equation}
%KVLnote: MJt in summaries, C2Jt in data
is shown in Fig.~\ref{fig:MJofTfits}. As expected, $M^{\rm
  J}_{\alpha}(T_{\rm f})$ is increasing with increasing $T_{\rm f}$
and can be fitted by Maxwell-Boltzmann statistics (lines) with $E_{\rm
  A}^{\rm Si}=E_{\rm A}^{\rm O}=2.89$~eV. Activation energies
determined from the diffusion coefficient \cite{BKSTc} are considerably
higher: $E_{\rm A}^{\rm Si}=5.18$~eV and $E_{\rm A}^{\rm O}=4.66$~eV.

%-------------
%-------------

\subsection{Correlation of jumps and defects}

\begin{figure}[h]
%% Note: snapshot.eps=fig14.eps got cut, so all figs here shifted by -1
\includegraphics[width=3.1in]{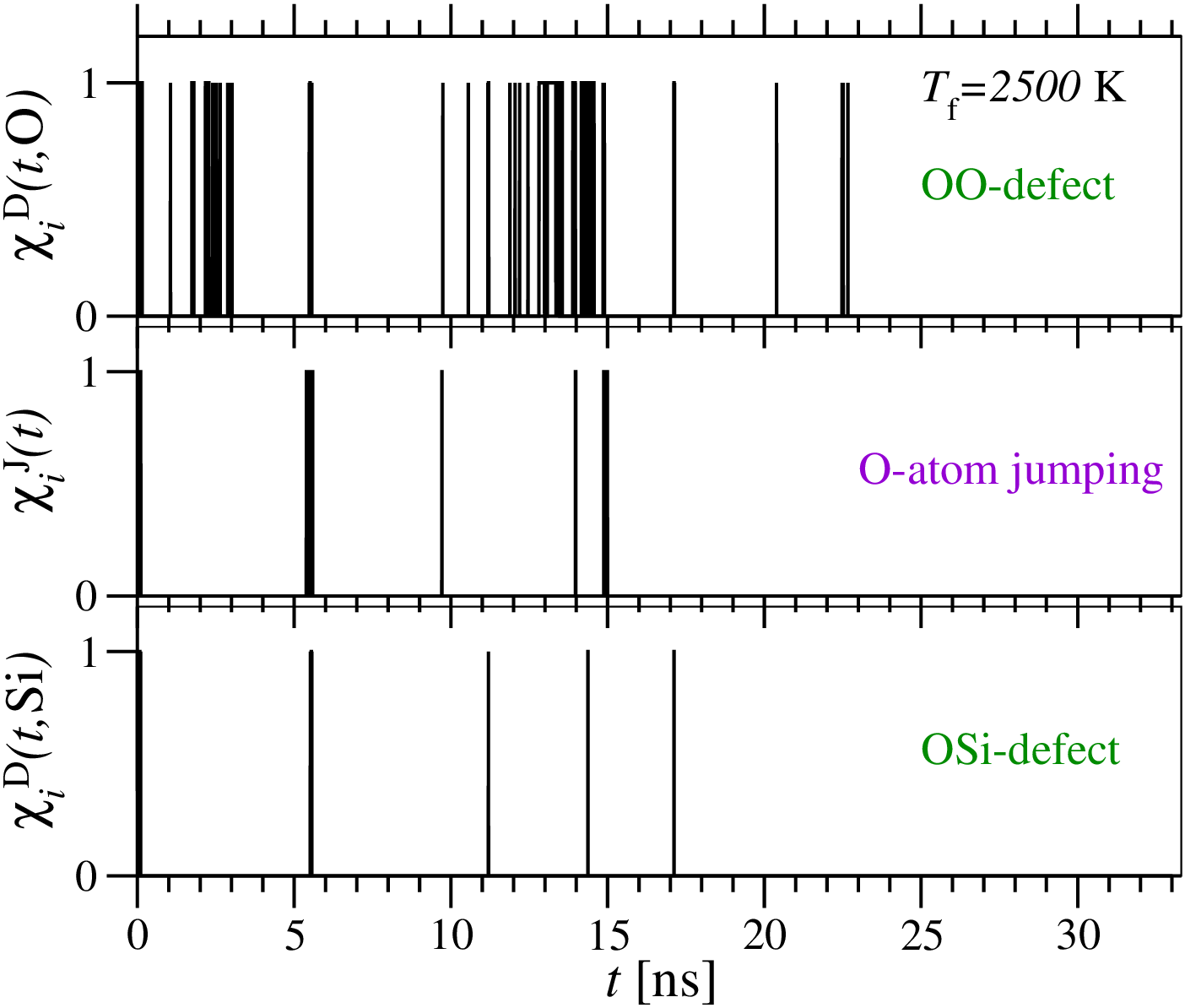}
\caption{\sf  For the comparison we show for an O-atom 
  the defect functions $\chi^{\rm D}_i(t)$ of OO-defects (top figure) 
 and of OSi-defects (bottom figure) and the jump function
$\chi^{\rm J}_i(t)$ (middle figure) 
  at $T_{\rm f}=2500$~K.
        }
\label{fig:chiDJ_2500from5000c08p134}
\end{figure}

So far we have investigated the
dynamics of the system from two perspectives: defects and jumps.  
Having identified jump events by $\chi^{\rm J}_i(t_l)$ and defects by
$\chi^{\rm D}_i(t_l,\beta)$, we can now quantify the correlations
between the two sorts of events. To illustrate our approach, we show
in Fig.~\ref{fig:chiDJ_2500from5000c08p134} $\chi^{\rm J}_i(t_l)$ and
$\chi^{\rm D}_i(t_l,\beta)$ for an O jumper whose trajectory is shown
in Fig.~\ref{fig:traj_from5000c08p134}. A similar plot for an Si
jumper is shown in Fig.~\ref{fig:chiDJ_2500from5000c08p134}. If the
horizontal lines of $\chi^{\rm D}_i(t,\beta)$ and $\chi^{\rm J}_i(t)$
are aligned, this implies a strong correlation of jumpers
and defects.

\vspace*{3mm}

%-------------
\begin{figure}[h]
\includegraphics[width=3.1in]{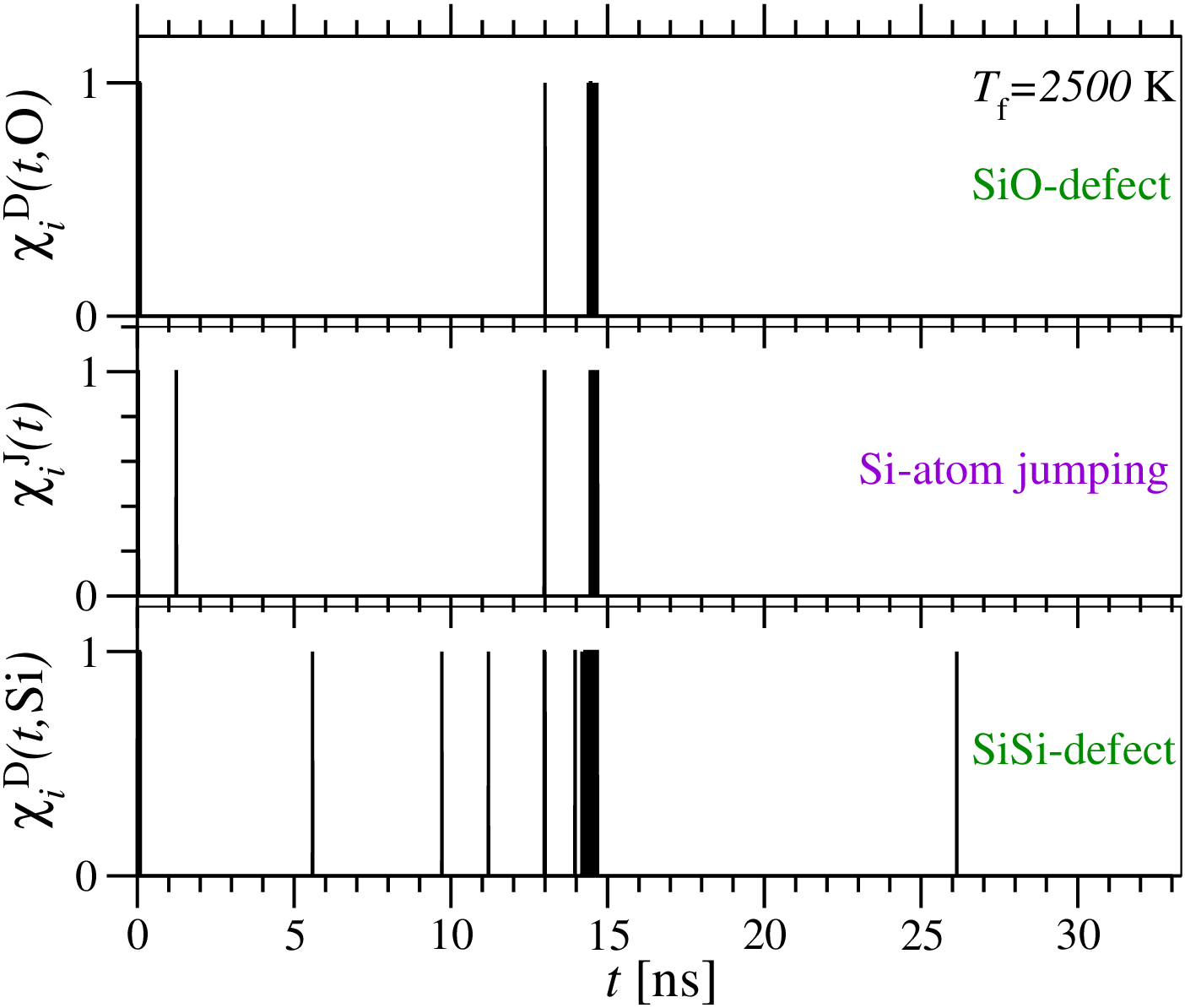}
\caption{\sf Similar to Fig.~\ref{fig:chiDJ_2500from5000c08p134} 
  also at $T_{\rm f}=2500$~K but for an Si-atom we show 
  the defect functions $\chi^{\rm D}_i(t)$ of SiO-defects (top figure)
 and of SiSi-defects (bottom figure) and the jump function
$\chi^{\rm J}_i(t)$ (middle figure). 
        }
\label{fig:chiDJ_2500from5000c08p56}
\end{figure}
%-------------

Both jumping atoms as well as defects are rare events at low
temperatures.  If atoms were jumping independently from creating
defects then the joint probability that an atom is a jumper and
simultaneously a defect would be given by the product of two very
small probabilities, which can be estimated as follows. The probability
$p$ for a particle to be a jumper can be approximated by the fraction
of jumpers $p=M^J_{\alpha}$ and similarly the probability $q$ for a
particle to be a defect by $q=M^D_{\alpha,\beta}$. If the events were
independent, the joint probability for a particle to be a defect and
%vs9:
%simultaneously a jumper is $pq$ 
%which is $2\times 10^{-5}/3\times 10^{-6}/2\times 10^{-6}/10^{-4}$ for 
%SiSi/SiO/OSi/OO defects at $T_f=2500$K, whereas
%Figs. \ref{fig:chiDJ_2500from5000c08p134} and \ref{fig:chiDJ_2500from5000c08p56}
%suggest a strong correlation.
%v11
simultaneously a jumper is $pq$. At $T_f=2500$ K the smallest
value of $pq$ is $2\times 10^{-6}$ for OSi defects and 
the largest value of $pq$ is $10^{-4}$ for OO defects. In contrast 
Figs. \ref{fig:chiDJ_2500from5000c08p134} and 
\ref{fig:chiDJ_2500from5000c08p56} indicate a larger likelihood of
$\chi^{\rm D}_i$ and $\chi^{\rm J}_i$ being aligned and thus
suggest a strong correlation.

A quantitative measure for the correlations of jumpers and defects is the
following correlation function:

\begin{widetext}
\begin{equation}
A^{\rm DJ}_{\alpha,\beta}  =\frac{ 
     \left \langle 
        \frac{1}{N_{\alpha}}
        \sum \limits_{i=1}^{N_{\alpha}}
            \chi^{\rm D}_i(t_l,\beta) \chi^{\rm J}_i(t_l)\right\rangle
-\left \langle 
        \frac{1}{N_{\alpha}}
        \sum \limits_{i=1}^{N_{\alpha}}
            \chi^{\rm D}_i(t_l,\beta) \right\rangle
\left \langle 
        \frac{1}{N_{\alpha}}
        \sum \limits_{i=1}^{N_{\alpha}}
            \chi^{\rm J}_i(t_l) \right\rangle}
{ 
     \left \langle        
        \frac{1}{N_{\alpha}}
            \sum \limits_{i=1}^{N_{\alpha}}
            \chi^{\rm D}_i(t_l,\beta)\right\rangle 
 \left \langle        
        \frac{1}{N_{\alpha}}
            \sum \limits_{i=1}^{N_{\alpha}}
            \chi^{\rm J}_i(t_l)\right\rangle 
  }
  \mbox{.}
%
%= \frac{
%     \left \langle 
%        \frac{1}{N_{\alpha}}
%        \sum \limits_{i=1}^{N_{\alpha}}
%            \chi^{\rm D}_i(t_l,\beta) \chi^{\rm J}_i(t_l)\right\rangle
%       }{M^{\rm D}_{\alpha \beta} M^{\rm J}_{\alpha}} - 1
 \label{eq:ADJ}
\end{equation}
\end{widetext}
%-------------

\begin{figure}[h]
\includegraphics[width=0.45\textwidth]{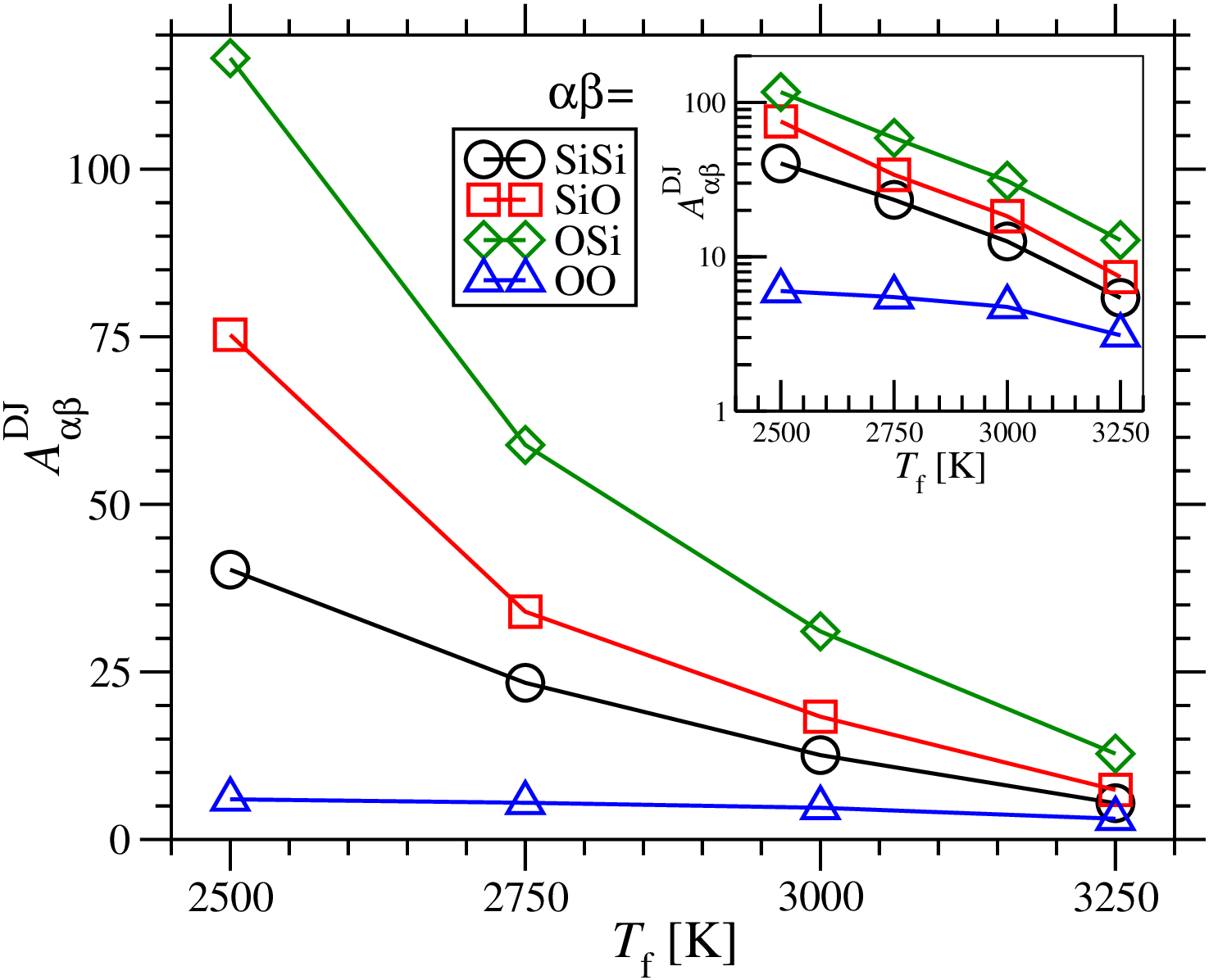}\\
\vspace{0.4cm}
\caption{\sf Correlation of defects and jumps $A^{\rm DJ}_{\alpha \beta}$
     as defined in Eq.~(\ref{eq:ADJ}). The lines are a guide to the eye.
 The inset shows the same data in a semilogarithmic plot.
             }
\label{fig:ADJ}
\end{figure}

The defect-jumper correlation is shown in Fig.~\ref{fig:ADJ}
for both types of jumpers (indicated by the first letters) and both
types of corresponding defects (indicated by the last letters). The
correlation of a defect involving a wrong coordination between an 
Si-atom and an O-atom is very high at low temperatures for both 
Si-atoms and O-atoms jumping (red and green line respectively). Only
the correlation between an O-atom which is jumping and not correctly
coordinated with other O-atoms is less well pronounced. 
We interpret these results as follows: A breakup of the
SiO$_4$ tetrahedra, which destroys the appropriate coordination
between Si and O-atoms, is likely to involve a jump, whereas jumps of
an O-atom, involving the motion of two tetrahedra with respect to
each other, can happen without the creation of a defect. 

\begin{figure}[h]
\includegraphics[width=3.1in]{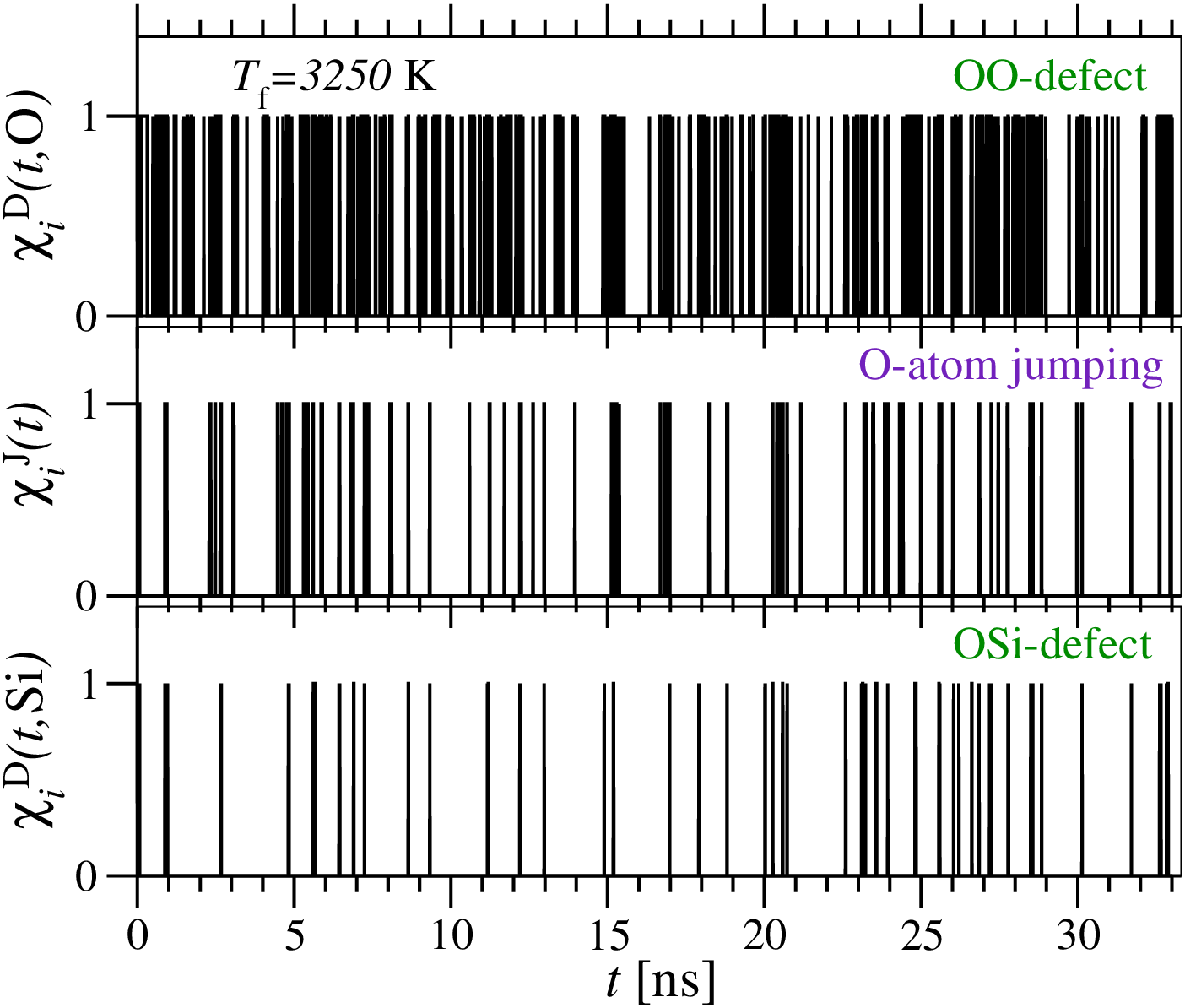}
\caption{\sf Similar to Fig.~\ref{fig:chiDJ_2500from5000c08p134} 
  for the same O-atom but here for $T_{\rm f}=3250$~K we show 
  the defect functions $\chi^{\rm D}_i(t)$ of OO-defects (top figure)
 and of OSi-defects (bottom figure) and the jump function
$\chi^{\rm J}_i(t)$ (middle figure). 
        }
\label{fig:chiDJ_3250from5000c08p134}
\end{figure}

At high temperatures the correlation of jumpers and defects is small, 
which can also be guessed from single trajectories, see
Fig. \ref{fig:chiDJ_3250from5000c08p134}. It is also apparent from the
single trajectory (Fig. \ref{fig:chiDJ_3250from5000c08p134}) that the
O-atom which jumps is very often simultaneously an OSi defect, whereas
OO defects are so frequent that hardly any correlation can be
detected.  This is reflected in the average correlation (see
Fig.~\ref{fig:ADJ}) of the O-jumper which is four times higher for OSi
defects than for OO-defects at $T=3250$K.

%We find that the temperature dependence of $A^{\rm DJ}_{\alpha \beta}(T_f)$ is
%approximately determined by $M^{\rm D}_{\alpha \beta}$ since 
%$A^{\rm DJ}_{\alpha \beta}(T_f) \approx 
% \left (\frac{B_{\alpha \beta}}{M^{\rm D}_{\alpha \beta}(T_f)}-1 \right )$
%as shown in Fig.~\ref{fig:ADJ}.

%-------------
%-------------

%The exact values of $A^{\rm DJ}_{\alpha \beta}$ 
%depend on details  of the defect and jump definitions: 
%When in Eq.~({eq:jumpdef}) 
%the factor $\sqrt{2}$ instead of $3$ is used, more jumps 
%are detected, i.e. the denominator is increased, 
%which leads to smaller values of 
%$A^{\rm DJ}_{\alpha \beta}$.  Similarly 
%for larger $r_{\rm min}^{\alpha \beta}$ we find more defects 
%and $A^{\rm DJ}_{\alpha \beta}$ decreases.
%However, the qualitative result is the same for all variations:
%We find a strong correlation between 
%single particle jumps and defects; the highest correlation is
%for SiO- and OSi-defects and for the lowest temperature.

%\subsection{Spatial Correlation of Defect- and Jump-Particles}
%
%{\bf Maybe we should comment on that for $i=j$ included, the 
%  contribution for $r=0$ is about 1000 times larger than peaks 
%  for $r>0$ (see summary I Fig.37 and Fig.38)}\\

%-------------
\section{Summary and Conclusions}
\label{sec:summary}

%In previous work \cite{kvl2010,kvl2013} we had
%performed molecular dynamics simulations 
%using the BKS-potential to model SiO$_2$ \cite{beest_90}.
%The system was quenched from high temperature 
%$T_{\rm i} \in \{3760$~K, $5000$~K$\}$ to 
%temperatures 
%$T_{\rm f} \in \{2500$~K, $2750$~K, $3000$~K, $3250$~K$\}$  
%which are below the glass transition $T_{\rm c}=3330$~K
%and at which the system behaves like a strong glass former.

In this paper we analyzed {\em time-averaged single particle 
trajectories} $\overline{\mathbf r}_i(t_l)$ at temperatures well below the glass transition temperature.
%$T_{\rm f}$.
%The time average filters out thermal vibrations, similar
%to inherent structures 
%\cite{Stillinger1984,Sastry1999,saksaengwijit2004,reinisch2005,saksaengwijit2006,saksaengwijit2007}. However, in our analysis we use directly 
%the time averaged trajectories, i.e. stay in real space.  
Loosely spoken, the time average allowed us to 
watch a movie of the complicated particle dynamics by filtering 
out the background noise of vibrations, 
revealing the underlying major relaxational processes.
Using $\overline{\mathbf r}_i(t_l)$ we determined the 
radial distribution function and coordination number distribution.
Both are very sharply peaked, reflecting a highly structured 
network of O-corner sharing SiO$_4$-tetrahedra  
in which almost all particles have the ideal coordination number.
%of type $\alpha \in \{$Si,O$\}$ are surrounded by
%$z_{\rm perfect}^{\alpha \beta}$ neighbors of type $\beta$   
%($z^{\rm SiSi}_{\rm perfect}=4$, $z^{\rm Si0}_{\rm perfect}=4$,
% $z^{\rm OSi}_{\rm perfect}=2$ and $z^{\rm OO}_{\rm perfect}=6$).
This led us to focus on deviations from this well-defined local
neighborhood in order to find the excitations which are responsible
for the slow structural relaxation in the random network.
 
We defined defects in the time-averaged structure by an indicator
function: $\chi_i^D(t_l,\beta)=1$, if particle $i$ of type $\alpha$ at
time $t_l$ has coordination
%  an $\alpha \beta$-defect event to occur for 
%particle $i$ at time $t_l$ when 
$z_i^{\alpha \beta}(t_l) \ne z_{\rm perfect}^{\alpha \beta}$.  We
computed the average number of defects and the time-delayed
autocorrelation of $\chi_i^D(t_l,\beta)$, from which we extracted the
average lifetimes of defects.  We observe a very strong variation of
lifetimes for different sorts of defects.  SiO- and OSi-defect
correlations decay fast; in the movie analogy they correspond to short
flashes which come and go. (Note, however, that we are looking at
time scales which are larger than jump times and huge as
compared to oscillation periods.) In contrast, OO-defects are very
long lived; their lifetime becomes comparable to the equilibration
time for $T=2750$ K. All lifetimes are strongly temperature dependent. 
E.g. the lifetime of OO-defects increases by a factor of $100$ in the
temperature range $2500$ K$ \le T \le 3250$ K. 
Given the rather mild temperature
dependence of the average structure as described by the pair
correlation, we expect that defects are one of possibly other
excitations which determine the temperature dependence of glassy
properties at low temperatures.  This issue needs to be explored
further in future work.

Local structural rearrangements are achieved by jumping atoms,
i.e. atoms which move considerably further than a typical oscillation
amplitude. The statistics of jumpers in SiO$_2$ has been studied in
previous work \cite{kvl2010,kvl2013}. We expect that single particle
jump events go hand in hand with the creation and annihilation of
defects and hence we have computed the correlation of defect and jump
events.  At low temperatures the correlation is very strong, in
particular for defects involving one Si- and one O-atom, emphasizing
the important role of defects for structural rearrangements.

It would be interesting to investigate spatial correlations of
defect and jump events and relate our work to studies of dynamic
heterogeneities in which the most mobile particles are selected 
(for reviews we refer the reader to \cite{Ediger2000,biroli2013}).  Spatial
correlations, however, require a larger simulation size, which is
planned for the future. Another interesting extension are
other network formers such as B$_2$O$_3$ and BeF$_2$, 
for which we expect defects to be well defined. We leave it for future work
to study their defect dynamics as it is presented here.
Our analysis of time-averaged trajectories shines light 
on the main features of structural changes and
is easily applicable to simulations and experiments of 
other network-forming and non-network forming systems 
and other strong and fragile glass formers.
It remains to be seen whether the defect and jump dynamics
presented here is a universal phenomenon.

\begin{acknowledgments}
KVL was supported by the Deutsche Forschungsgemeinschaft
via SFB 602 and FOR1394. KVL thanks 
the Institute of Theoretical Physics, University of G{\"{o}}ttingen,
for financial support and hospitality. We thank B. Vollmayr-Lee for 
fruitful discussions and comments on an earlier version of this manuscript. 
\end{acknowledgments}

\bibliography{SiO2defects_v11}% Produces the bibliography via BibTeX

\end{document}